\documentclass[conference]{IEEEtran}
\IEEEoverridecommandlockouts

\usepackage{cite}
\usepackage{amsmath,amssymb,amsfonts}
\usepackage{algorithmic}
\usepackage{graphicx}
\usepackage{textcomp}
\usepackage{xcolor}
\usepackage{hyperref}

\usepackage{array}
\usepackage{makecell}
\usepackage{multirow}
\usepackage{algorithm}
\usepackage{subfigure}
\usepackage{color}
\usepackage{booktabs}

\def\BibTeX{{\rm B\kern-.05em{\sc i\kern-.025em b}\kern-.08em
    T\kern-.1667em\lower.7ex\hbox{E}\kern-.125emX}}
\begin{document}

\title{

LOVO: Efficient Complex Object Query in Large-Scale Video Datasets
\\

\author{
\IEEEauthorblockN{
    Yuxin Liu\IEEEauthorrefmark{3}, 
    Yuezhang Peng\IEEEauthorrefmark{3},
    Hefeng Zhou\IEEEauthorrefmark{3}, 
    Hongze Liu\IEEEauthorrefmark{3},
    Xinyu Lu\IEEEauthorrefmark{3}, \\
    Jiong Lou\IEEEauthorrefmark{3}\IEEEauthorrefmark{1}, 
    Chentao Wu\IEEEauthorrefmark{3},
    Wei Zhao\IEEEauthorrefmark{2},
    Jie Li\IEEEauthorrefmark{3}\IEEEauthorrefmark{1}
    }

\IEEEauthorblockA{\IEEEauthorrefmark{3}\textit{Shanghai Jiao Tong University}, China}
 \IEEEauthorblockA{\IEEEauthorrefmark{2}\textit{Shenzhen University of Advanced Technology}, China\\ \{emotion-xin, lj1994, lijiecs\}@sjtu.edu.cn}
 
}

\thanks{
  
  Corresponding authors are Jie Li\IEEEauthorrefmark{1} and Jiong Lou\IEEEauthorrefmark{1}.

}

}

\maketitle

\begin{abstract}
The widespread deployment of cameras has led to an exponential increase in video data, creating vast opportunities for applications such as traffic management and crime surveillance. However, querying specific objects from large-scale video datasets presents challenges, including (1) processing massive and continuously growing data volumes, (2) supporting complex query requirements, and (3) ensuring low-latency execution. Existing video analysis methods struggle with either limited adaptability to unseen object classes or suffer from high query latency. 

In this paper, we present LOVO, a novel system designed to efficiently handle comp\underline{L}ex \underline{O}bject queries in large-scale \underline{V}ide\underline{O} datasets. Agnostic to user queries, LOVO performs one-time feature extraction using pre-trained visual encoders, generating compact visual embeddings for key frames to build an efficient index. These visual embeddings, along with associated bounding boxes, are organized in an inverted multi-index structure within a vector database, which supports queries for any objects. During the query phase, LOVO transforms object queries to query embeddings and conducts fast approximate nearest-neighbor searches on the visual embeddings. Finally, a cross-modal rerank is performed to refine the results by fusing visual features with detailed textual features. Evaluation on real-world video datasets demonstrates that LOVO outperforms existing methods in handling complex queries, with near-optimal query accuracy and up to 85x lower search latency, while significantly reducing index construction costs. This system redefines the state-of-the-art object query approaches in video analysis, setting a new benchmark for complex object queries with a novel, scalable, and efficient approach that excels in dynamic environments.

\end{abstract}

\section{Introduction} \label{sec:intro}

The proliferation of cameras in public spaces and on mobile devices has led to an explosion of video data~\cite{cameras,focus,dovedb,exsample,noscope}. The global market for surveillance cameras has been expanding steadily, with the number of cameras worldwide reaching 1 billion in 2021~\cite{number}. A camera, capturing video at 30 frames per second, can generate up to 20 GB of data per day~\cite{zilla}. Considering the large number of cameras, the video data generated every day is staggering. This massive volume of video data presents valuable opportunities in many areas, such as urban governance~\cite{trafficMonitor}, traffic flow optimization~\cite{trafficControl}, crime tracking~\cite{prevention}, and emergency response monitoring~\cite{emergency}. These tasks often rely on complex object-centric querying to identify and track specific targets within dynamic environments~\cite{otif,blazeit,seiden,eva,everest,figo,tahoma,leap}. Object query forms the foundation of these video analysis applications, so it is urgent to design an efficient system for complex object queries within large-scale video datasets. 

\begin{figure}[t]
    \centering
    \subfigure[Query-agnostic index-based methods (QA-index).]{
        \includegraphics[width=\linewidth]{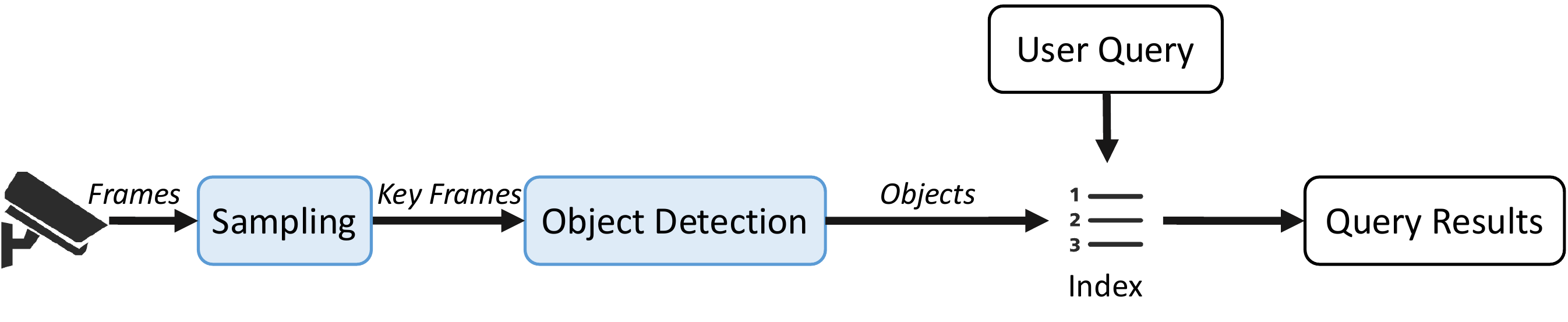}
        \label{fig:intro1_1}
    }
    \subfigure[Query-dependent search-based methods (QD-search).]{
        \includegraphics[width=\linewidth]{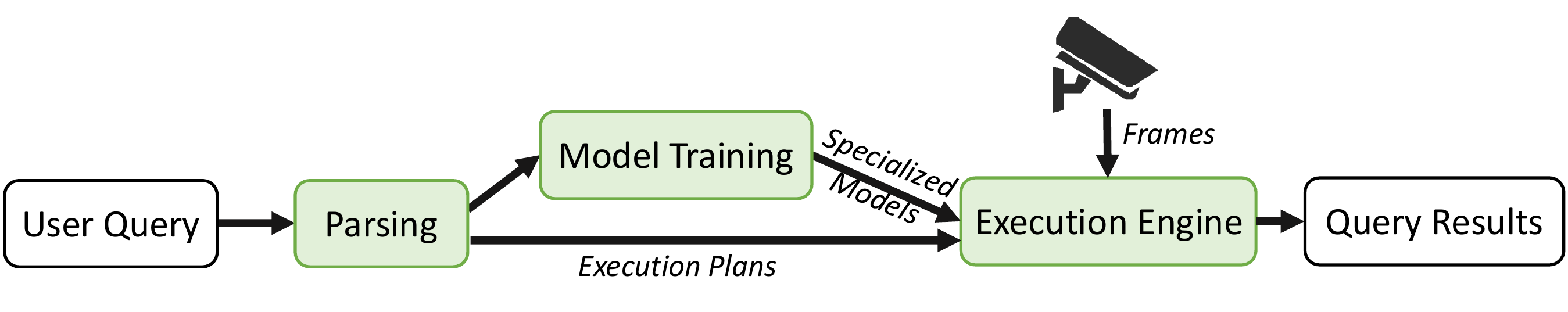}
        \label{fig:intro1_2}
    }
    \subfigure[LOVO.]{
        \includegraphics[width=\linewidth]{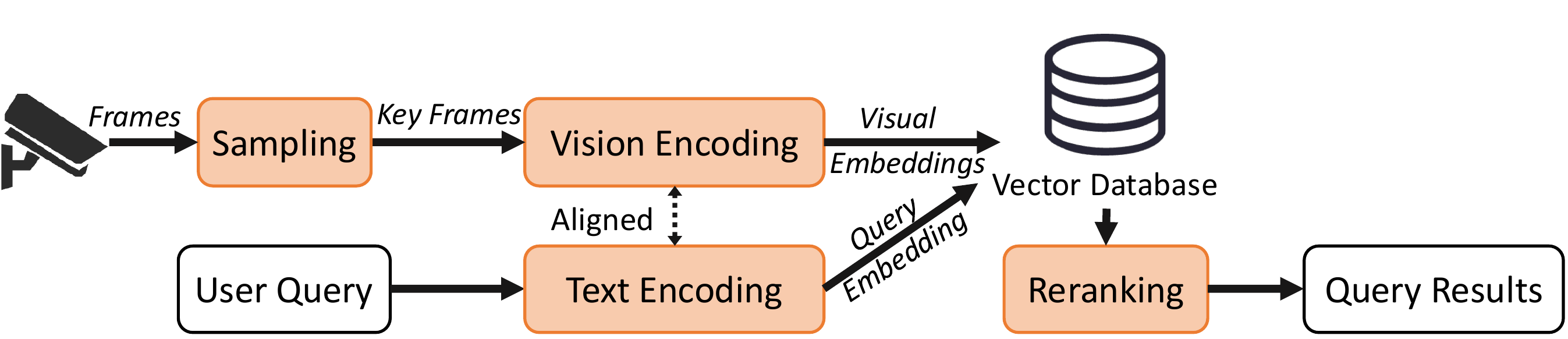}
        \label{fig:intro1_3}
    }
    \vskip -0.03in
    \caption{Existing methods vs. LOVO.}\label{fig:intro1}
    \vspace{-15pt}
\end{figure}

Querying specific objects from massive video datasets is nontrivial, presenting several significant challenges. (1) \textit{Massive and continuously growing video data}. It is very expensive to query a specific object in such large and growing video datasets. (2) \textit{Complex query requirements}. To support fine-grained complex queries for arbitrary objects, video retrieval systems must accommodate object queries that go beyond the predefined classes in traditional object detection~\cite{coco} and incorporate more detailed descriptions based on natural human language. (3) \textit{Low-latency query execution}. Achieving low-latency queries is essential to improve user experience and satisfy the requirements of latency-sensitive video analysis applications. However, this becomes more difficult when faced with the dual challenge of large-scale data and complex query specifications, where minimizing latency without compromising accuracy is a significant hurdle.

Existing object query methods in large-scale video datasets can be mainly classified into two categories, as shown in Fig.~\ref{fig:intro1}: Query-agnostic index-based methods (QA-index)~\cite{otif,equi, seiden,leap,zilla,chen2022ranked} and Query-dependent search-based methods (QD-search)~\cite{tahoma,noscope,blazeit,PP,dovedb,figo,miris,anderson2019physical}. Before receiving the object queries, QA-index methods extract objects of predefined classes from sampled video frames to build a simple index, so that the future query will be responded to with low latency. However, these methods cannot handle queries for objects out of predefined classes or with detailed descriptions, requiring significant time to retrain the object detection model for unseen classes. In contrast, QD-search methods parse each user query and generate customized execution plans, tailored to individual needs. This process involves training object detection models and applying them cooperatively. This provides enhanced flexibility to handle complex queries. However, QD-search methods suffer from inefficiency, as they repeatedly process the same video data for every distinct query, leading to significant latency.

To solve the above challenges, we design LOVO, an efficient system for complex object queries in large-scale video dataset, shown in Fig.~\ref{fig:intro1_3}. Leveraging the decoupled object detection and text encoder design, LOVO eliminates the need for repetitive video processing by performing one-time feature extraction offline, transforming key frames into visual embeddings stored in an inverted-indexing vector database. By aligned query embedding, LOVO transcends the constraints of predefined classes and supports detailed object descriptions. This system efficiently identifies candidate objects in rapid similarity computations via approximate nearest-neighbor searches and refines results through cross-modality rerank.

The system's capabilities are realized via three modules: (1) \textbf{Video Summary}. This module analyzes raw video data by extracting key frames and converting them into collections of semantic feature vectors with bounding box coordinates. Its decoupled encoder supports one-time video processing without predefined classes and text queries, enabling complex object localization. (2) \textbf{Database Storage}. This module manages the storage of object-level vectors using product quantization and inverted index mechanism, and also enables fast search via approximate nearest neighbor search. It ensures efficient organization and rapid query response, making it suitable for large-scale video dataset. (3) \textbf{Query Strategy}. This module employs a two-stage strategy: \textit{fast search} quickly identifies top-\textit{k} relevant object based on similarity to the query, while \textit{cross-modality rerank} refines the results by fusing text queries with the object’s visual information. This method enhances the accuracy of object queries in large video dataset with low latency. With an orthogonal design, LOVO allows flexible substitution of keyframe extraction algorithms, visual backbone models, and indexing methods tailored to specific needs.

To our knowledge, LOVO is the first complex object query system for large-scale video dataset, surpassing prior systems in supported workloads. Our contributions are as follows:

\begin{itemize}
    \item We identify the key challenges in implementing complex object query for large-scale videos, including high computational cost, limited query flexibility, and scalability constraints in open-world settings.
    \item We present LOVO, a novel system for large-scale video analysis. Its decoupled encoder transforms raw videos into a one-time semantic feature index in a vector database, free from predefined classes or text queries. With a two-stage query strategy, it enables low-latency, complex object queries for real-world use.
    \item We demonstrate the deployment of LOVO within a video database management system for complex object query in large-scale video data. Our system achieves both fast query response times and high retrieval accuracy, even in large-scale video environments.
    \item We conducted extensive experiments comparing LOVO with baseline systems on real-world video datasets. Results demonstrate that LOVO achieves near-optimal performance in complex object query. 
\end{itemize}

Section~\ref{sec:pre} discusses our motivation. Section~\ref{sec:overview} provides an overview of LOVO and introduces the rationale behind it. Section~\ref{sec:summary} details video processing. Section~\ref{sec:storage} covers the indexing construction within the video database. Section~\ref{sec:query} proposes query strategies. Section~\ref{sec:eval} presents and analyzes experimental results, Section~\ref{sec:related} reviews related work, and Section~\ref{sec:conclu} concludes the paper and discusses future work.
\section{Motivation}\label{sec:pre}
In this section, we present the motivation behind our design through illustrative experiments.

\begin{figure}[t]
    \centering
    \subfigure[Execution time for existing methods across three query complexities.]{
        \includegraphics[width=0.9\linewidth]{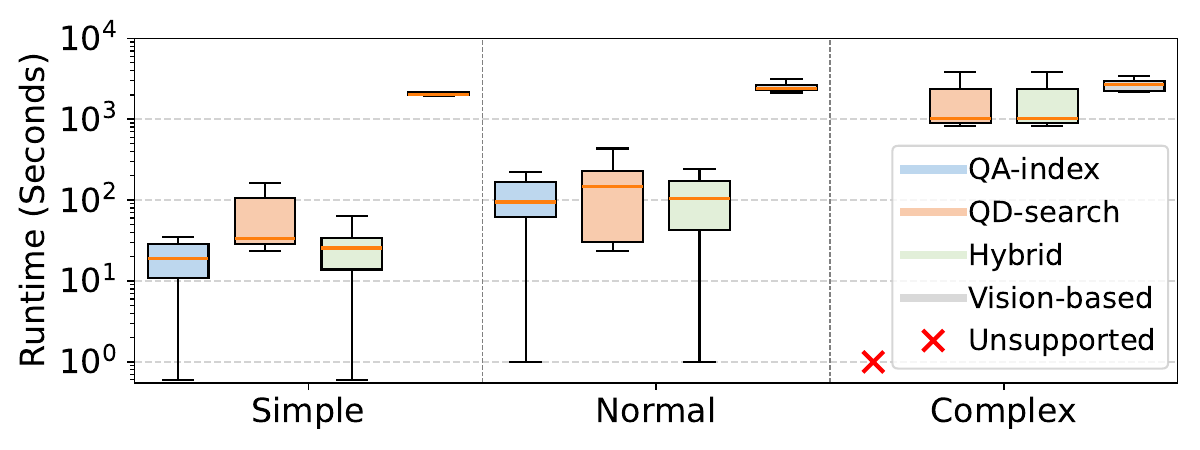}
        \label{fig:pre1_1}
    }
    \subfigure[Qualitative performance comparison of object query methods. QA-index method is capable of handling queries for predefined classes, while QD-search is suited for queries with novel features. Vision-based, on the other hand, fully supports complex natural language queries.]{
        \includegraphics[width=0.96\linewidth]{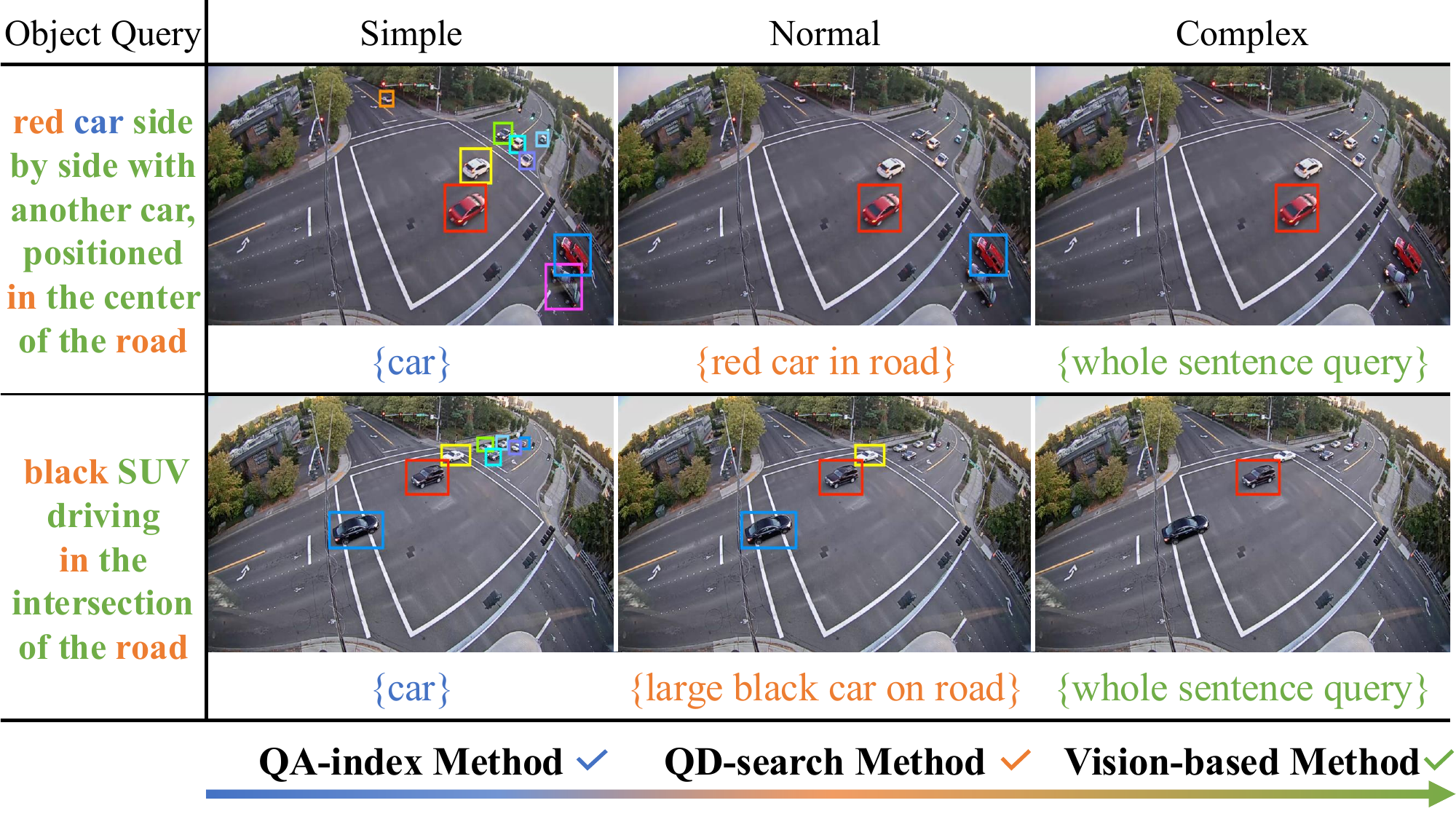}
        \label{fig:pre1_2}
    }
    \caption{Efficiency and performance comparison of existing methods.}\label{fig:pre1}
    \vspace{-10pt}
\end{figure}

\textbf{Experimental Design.} We compare existing object query methods in video analysis, including QA-index, QD-search, hybrid methods that combine both, and vision-based methods leveraging large vision-language models. For the QA-index method, we choose OTIF~\cite{otif} and VOCAL~\cite{equi}. For QD-search, we focus on FIGO~\cite{figo} and MIRIS~\cite{miris}. For vision-based methods, we choose DINO~\cite{dino} for its generalization ability for complex queries. We focus on measuring execution time because video processing times are typically imperceptible to users. To ensure the fairness of the experiment, we do not consider additional model retraining for the methods. Fig.~\ref{fig:pre1_1} illustrates execution time taken by each method for a single query, while Fig.~\ref{fig:pre1_2} shows performance of methods.

\textbf{Query Complexity}. The experiments are conducted on a real-world surveillance video dataset collected from an intersection in Bellevue~\cite{bellevue}. To evaluate the methods' performance for object queries, we use two representative queries shown in Fig.~\ref{fig:pre1_2}. To evaluate the capability of the existing methods, we decompose the queries into three levels of complexity and test each method with all three types of queries: a simple query (``car'' from MSCOCO labels~\cite{coco}), normal queries (``red car in road'' and ``large black car on road'' with novel features), and a complex query involving the whole sentence (since ``SUV" is a new class relative to the traditional ``car," ``side by side with another car, both positioned in the center of the road"and ``driving in the intersection of the road" are detailed descriptions of the object's behavior and location).

\textbf{QA-index Methods}. These methods excel in handling predefined queries by leveraging key frame extraction and object detection to build static indexes. They are adept at recognizing common objects, such as ``car", within fixed classes, as these detection models are trained on established label sets like MSCOCO. The execution times for these systems are quite efficient, around $0.5$ seconds. However, QA-index methods are fundamentally limited in handling complex queries, such as acronyms like ``SUV" or detailed descriptions in a sentence, which do not directly map to their predefined classes. Consequently, their pre-built indexes cannot accommodate nuanced descriptions or unseen object categories, significantly limiting their applicability in open-world environments.

\textbf{QD-search Methods}. These methods offer flexibility by applying multiple specialized detection models based on the user query. This allows QD-search methods to handle normal, descriptive queries with novel features, such as ``red car in the road" or ``large black car on the road". However, this flexibility comes at the cost of computational efficiency. Because QD-search methods operate by using models to traverse a large portion of the video dataset in response to each query, they become computationally infeasible in large-scale scenarios. Moreover, these methods struggle with queries that require spatial relationships, such as ``side by side" or ``in the center," which necessitate a deeper semantic understanding compared to simpler terms like ``Near" or ``FrontOf," unless additional retraining is performed. Consequently, QD-search methods are less efficient for complex object querying.

\renewcommand{\arraystretch}{1.2} 
\begin{table}[t]
    \centering
    \caption{Comparison of Capabilities Across Object Query Methods}
    \label{tab:pre3}
    \begin{tabular}{l c c c}
    \hline
    \textbf{Capability} & \textbf{QA-index} &  \textbf{QD-search} & \textbf{Vision-based} \\ \hline
    Predefined Classes & Yes & Yes & Yes \\ \hline
    Simple Descriptions & No & Yes & Yes \\ \hline
    Complex Queries & No & No & Yes \\ \hline
    Scalability & Yes & Moderate & No \\ \hline
    Video Preprocessing & Extensive & Minimal & Moderate \\ \hline
    Execution Efficiency & High & Low & Low \\ \hline
    Query Accuracy & High & Moderate & High \\ \hline
    \end{tabular}
    \vskip -0.1in
\end{table}

\begin{figure*}[ht]
    \begin{center}
     \includegraphics[width=\textwidth]{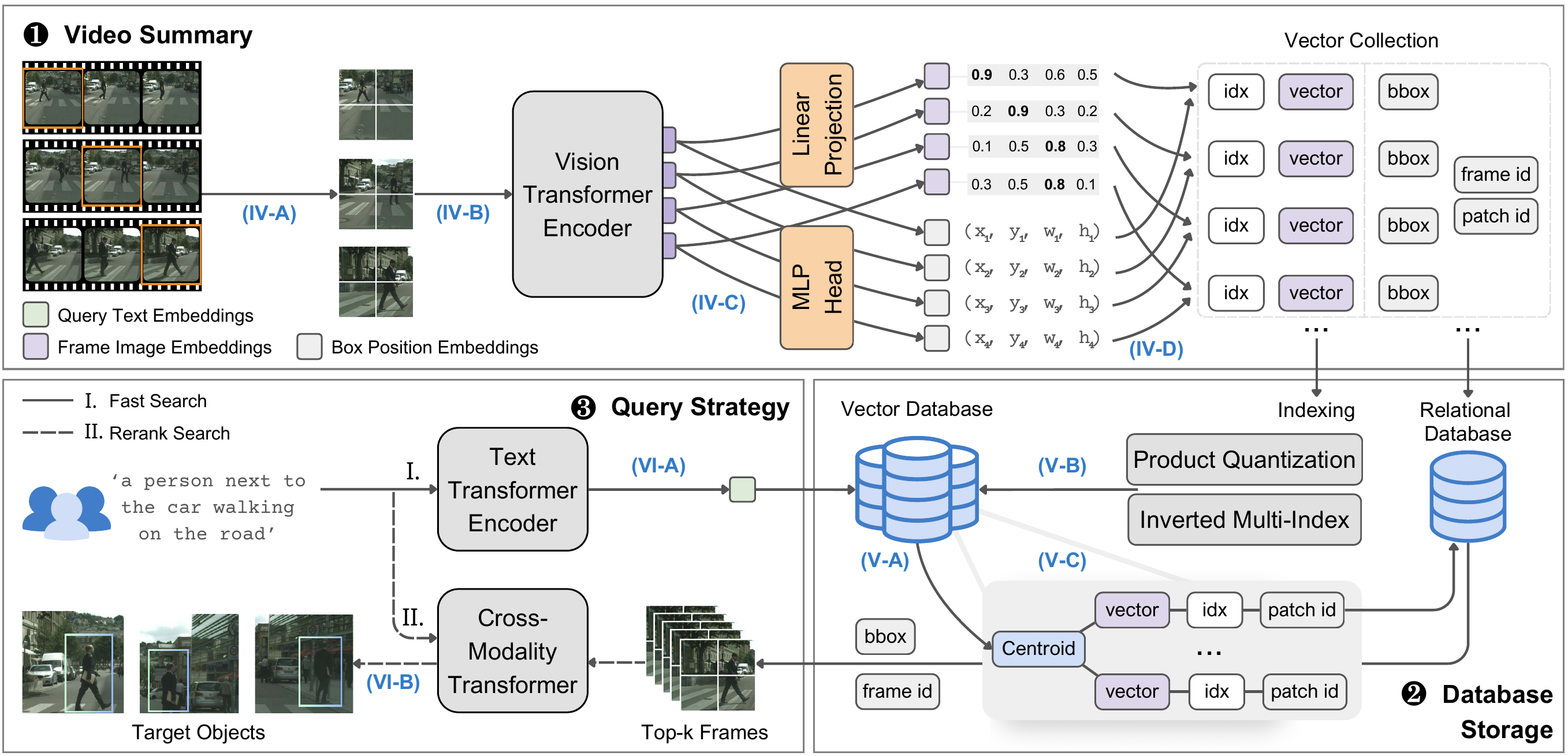}
     \vskip -0.05in
    \caption{LOVO workflow for complex object query in large-scale video database.}
    \label{fig:workflow}
    \vspace{-10pt}
    \end{center}
\end{figure*}

\textbf{Hybrid Methods}. These attempts aim to bridge the gap between QA-index and QD-search methods by combining static indexing with on-demand search. In scenarios where the pre-built index successfully identifies a category, hybrid methods benefit from the indexing efficiency. However, when the index fails, these methods revert to the QD-search approach, incurring substantial computational overhead due to redundant video scanning. Consequently, the performance gains from combining indexing and searching are minimal, and these methods do not significantly outperform individual QA-index or QD-search methods. Due to these inefficiencies, hybrid methods are excluded from further experimental comparisons.

\textbf{Vision-Based Methods}. Vision-based methods that utilize large-scale vision-language techniques, such as DINO~\cite{dino}, offer flexibility in processing complex, natural language queries, including entire sentence descriptions. Unlike QA-index methods, which require extensive preprocessing to detect objects in advance, these methods dynamically extract visual features relevant to the text query without depending on predefined classes. While Vision-based methods are capable of understanding detailed descriptions, they still involve high computational resource requirements and significant inference time. Consequently, although these models perform well in terms of accuracy and flexibility, their high costs and latency make them impractical in large-scale video analysis.

The aforementioned limitations of existing methods summarized in Table~\ref{tab:pre3}, underscore need for the system that enables low-latency complex object query in large-scale videos.

\section{LOVO}\label{sec:overview}
In this section, we introduce LOVO, a cutting-edge framework for efficient complex object queries in large-scale video datasets. To enable low-latency object queries, an index must first be pre-constructed, then the future object query will only involve simple index lookup~\cite{focus}. However, these indexes, including object classes, frame IDs, and limited spatial-temporal correlations, can hardly support complex object queries (with unseen object classes and detailed descriptions). The major difference between LOVO and previous QA-index methods is: LOVO utilizes visual embeddings instead of object classes to index video frames. Visual embeddings contain richer context information than object classes to support complex object queries. Then, the query process is converted to embedding similarity match and substantially speeded up by Approximate Nearest Neighbor Search. Finally, the query results are reranked by the Cross-Modelity Transformer, which utilizes a cross-attention layer between language and vision for feature enhancement. As illustrated in Fig.~\ref{fig:workflow}, LOVO comprises three key modules: Video Summary, Database Storage, and Query Strategy. Design rationale of modules is explained as follows.

\textbf{Video Summary}. This module processes raw videos by extracting key frames and converting them into semantic features using a decoupled visual encoder, then summarizes the object-level information into the bounding box coordinates. To tackle the challenge of massive, continuously updated video data, this module enables one-time feature extraction, transforming videos into compact, query-friendly object vectors ready for indexing in a database. This eliminates the need for repetitive processing, allowing for fast, on-demand queries without reprocessing the entire dataset. To support complex object queries, the module uses object localization in key frames, enabling flexible searches beyond the predefined classes. Additionally, visual and text encoders are decoupled, avoiding the computational overhead of early fusion. In this way, we enable fast and scalable searches when querying.

\textbf{Database Storage.} This module is designed to efficiently store object features, corresponding bounding boxes, and key frame IDs in a vector database, using vector quantization and inverted multi-index. These techniques enable scalable and high-performance storage of large video datasets while supporting fast query search via approximate nearest neighbor search. The primary goal of this module is to address the challenges of large-scale video data storage and low-latency query execution. By pre-indexing the feature vectors, the system eliminates the need for full database scans during query processing, significantly reducing query latency. This ensures fast responses to textual and object-based queries.

\textbf{Query Strategy.} This module adopts a two-stage approach: fast search and cross-modality rerank. In fast search, similarity matching is performed to recall a set of candidate objects. The rerank step then refines these candidates using fine-grained visual and textual features, ensuring precise object matching and more accurate results. The primary rationale behind this two-stage query strategy is to handle large-scale videos efficiently while minimizing latency. The fast search step performs quick similarity matching without the computational cost of deep text-visual fusion. Rerank narrows the search space and enhances accuracy by integrating cross-modality text-visual information for fine-grained object queries.

\section{Video Summary}\label{sec:summary}
This section explains how raw videos are transformed into semantic feature vectors and indexed in a single round. 

\subsection{Video Key frame Extraction}
To efficiently reduce data volume while preserving essential information, we represent each video as a sequence of key frames. A video collection \( V = \{v_i\}_{i=1}^M \) consists of \( M \) videos, with each video \( v_i \) composed of \( n_i \) key frames, denoted as \( v_i = \{ f^i_1, f^i_2, \dots, f^i_j, \dots, f^i_{n_i} \} \), where \( i \in \{1, \dots, M\} \) and \( j \in \{1, \dots, n_i\} \) are positive integers. Each frame \( f^i_j \in \mathbb{R}^{H \times W \times 3} \) is a high-resolution image with \( H \times W \) dimensions and 3 color channels (RGB). Instead of processing all individual frames, we focus on extracting key frames that effectively summarize the content of adjacent frames.

Key frame extraction combines temporal and content-based strategies. The temporal strategy selects frames at fixed intervals or scene changes, while the content-based strategy targets frames with notable visual differences or key events. We use the real-time tracking algorithm MVmed~\cite{mv}, which analyzes motion vectors to propagate detections across frames. Significant motion vector changes indicate scene shifts or high activity, making these frames ideal key frame candidates. This algorithm can be orthogonally adapted based on specific needs in various scenarios.

\subsection{Visual Patch Processing}
To capture object-level information and perform precise localization, each key frame \( f^i_j \) is divided into smaller patches. Specifically, the frame represented as a high-resolution image with dimensions \( H \times W \) and 3 color channels, is partitioned into \( K = \left\lfloor \frac{H}{S} \right\rfloor \times \left\lfloor \frac{W}{S} \right\rfloor\) patches, where \( K \in \mathbb{N}^* \) denotes the total number of patches, and \( k \in \{1, \dots, K\} \) are positive integers indexing the patches. Each patch is denoted as \( p_{jk} \in \mathbb{R}^{S \times S \times 3} \), representing a spatial region of size \( S \times S \) pixels extracted from the frame. To simplify the notation, the indexing of patches omits explicit references to the video \( i \). 

Inspired by the advancements in Owl-ViT~\cite{owlvit}, which transition from image-level classification to more granular object-level detection, we adapt this method to video key frame analysis by utilizing a Vision Transformer (ViT) as the image encoder. While Owl-ViT employs both a visual and a text encoder for cross-modality tasks, our method focuses solely on visual information during the video processing stage. Each patch is processed by the standard ViT encoder~\cite{vit} to extract visual features independent of textual queries, we avoid unnecessary computational overhead and enable single-pass feature extraction. The spatial information and semantic context are captured through the encoder’s multi-head attention mechanism, allowing each patch to learn from the surrounding patches’ features within the frame. 
Specifically, the ViT encoder transforms each patch \( p_{jk} \)  into a fixed-dimensional embedding \( \mathbf{z}_{jk} = \text{Encoder}(p_{jk}) \), where \( \mathbf{z}_{jk} \in \mathbb{R}^{D} \), and \( D \) denotes the embedding dimension (e.g., \( D = 768 \) for ViT-B/32). Instead of performing token pooling and final projection to aggregate these embeddings into a single global variable, we retain each patch’s individual embedding \( \mathbf{z}_{jk} \) to preserve its spatial and semantic details. 

\subsection{Object Localization}
\begin{figure}[t]
    \begin{center}
     \includegraphics[width=0.96\linewidth]{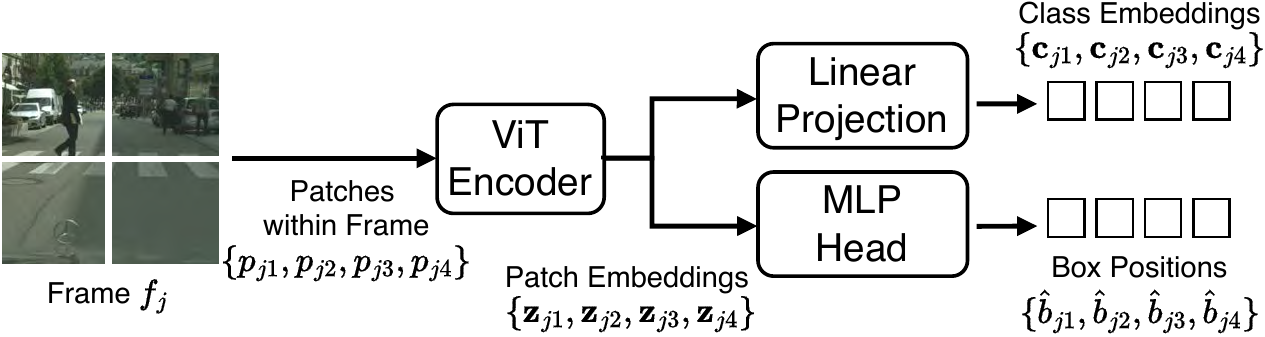}
     \vskip -0.05in
    \caption{Workflow of video summary process.}
    \label{fig:video_pre}
    \vspace{-10pt}
    \end{center}
\end{figure}

To meet the needs of object localization using complex query, we adapt the ViT architecture for object detection by removing the token pooling and final projection layers, which are typically used for image-level classification. Instead, lightweight object classification and localization heads are directly attached to the output tokens of the encoder. Each output patch token \( \mathbf{z}_{jk}\) represents a localized region of the image, allowing the model to predict object-level information for each token. The detailed workflow of this process is illustrated in Fig.~\ref{fig:video_pre}.

Once the embeddings for each patch are obtained from the patch processing, bounding box coordinates are predicted through a multi-layer perceptron (MLP). The predicted box for each patch \( p_{jk} \) is represented by the coordinates \(\hat{b}_{jk} \) by \( \hat{b}_{jk} = \text{MLP}(\mathbf{z}_{jk}) + b_{jk}^{\text{default}}\), where \( \mathbf{z}_{jk} \) is the visual embedding for the \( k \)-th patch in the \( j \)-th frame, and \( b_{jk}^{\text{default}} \) represents a default box. This default box, also referred to as an anchor box, is predefined based on the spatial layout of the corresponding patch and serves as an initial estimate of the bounding box location. MLP refines this estimate by predicting an offset that adjusts the anchor box to fit the object within the patch.

Additionally, the patch embeddings \( \mathbf{z}_{jk}\) are passed through a classification head. Here, the embeddings are projected into a lower-dimensional space \(D'\) to form the final class embeddings \(\mathbf{c}_{jk} \in \mathbb{R}^{D'}\), where \(D'\) is typically smaller than the original embedding dimension \(D\). This reduces the number of parameters required for downstream tasks and accelerates the process, ensuring low-latency inference.

This method supports open-vocabulary object localization, allowing the system to operate beyond predefined classes. Nevertheless, the limitation is that the relatively small patch size may hinder the accurate localization of objects spanning multiple patches, as larger cross-patch object features may be fragmented. Future work aims to improve cross-patch context aggregation to handle such cases better.

\subsection{Vector Collection Construction}
In preparation for efficient indexing in a vector database, for each key frame, we construct a collection that stores the key frame ID along with its corresponding \(K\) patch feature vectors and their associated bounding box coordinates. This collection can be formally represented as $\mathcal{I} = \{ (f_j, \{(\mathbf{c}_{j1}, \hat{b}_{j1}), (\mathbf{c}_{j2}, \hat{b}_{j2}), \dots, (\mathbf{c}_{jk}, \hat{b}_{jK}) \}) \}$, where \(f_j\) represents the \(j\)-th key frame, \(\mathbf{c}_{jk}\) is the visual embedding of the \(k\)-th patch in key frame \(f_j\), and \(\hat{b}_{jk}\) is the corresponding bounding box for patch \(k\).

\section{Database Storage}\label{sec:storage}
This section introduces features storage by similarity metrics, indexing, and approximate nearest neighbor search. 

\subsection{Similarity Metrics}
Similarity metrics are critical for identifying object features relevant to a given query. In our system, a textual query is encoded as a vector \( \mathbf{q} \), and its similarity to a vector \( \mathbf{c}_{jk} \) is computed using the dot product, prioritizing the embedding that maximizes \( \mathbf{q} \cdot \mathbf{c}_{jk} \). All vectors are normalized to unit \( L2 \)-norm, aligning the dot product directly with cosine similarity \( \text{cos}(\mathbf{q}, \mathbf{c}_{jk}) = \mathbf{q} \cdot \mathbf{c}_{jk} \). 

This normalization simplifies the relationship between similarity and distance metrics: a higher similarity implies a smaller Euclidean distance \( d(\mathbf{q}, \mathbf{c}_{jk}) = \sqrt{2 - 2 (\mathbf{q} \cdot \mathbf{c}_{jk})} \). As \( \mathbf{q} \cdot \mathbf{c}_{jk} \) approaches 1, the distance \( d(\mathbf{q}, \mathbf{c}_{jk}) \) approaches 0, indicating closer geometric alignment.

\subsection{Index Construction}
To achieve efficient storage and fast search response, we construct an index within a vector database, incorporating quantization techniques and an inverted multi-index structure.

To store class embeddings \( \mathbf{c}_{jk} \) with dimensionality \( D' \) in the vector database, we adopt Product Quantization (PQ), which decomposes the high-dimensional vector space into multiple subspaces~\cite{quanti}. Specifically, the embedding space \(\mathbb{R}^{D'}\) is divided into \( P \) subspaces, each with dimension \( m \), satisfying \( D' = P \cdot m \), where \( P, m \in \mathbb{N}^* \). Each embedding \( \mathbf{c}_{jk} \in \mathbb{R}^{D'} \) is expressed as a concatenation of \( P \) smaller components $\mathbf{c}_{jk} = \left([\mathbf{c}_{jk}]_1, [\mathbf{c}_{jk}]_2, \dots, [\mathbf{c}_{jk}]_P\right)^\top$,
where \([\mathbf{c}_{jk}]_p \in \mathbb{R}^m\) represents the \( p \)-th subspace component.

Each subspace of vectors is quantized independently into \( M \) clusters based on proximity to cluster centroids. This quantization maps each subspace vector to its nearest centroid:
\vspace{-10pt} 
\begin{align*}
    w_p : \mathbb{R}^m &\to \mathcal{C}_p, \\
    [\mathbf{c}_{jk}]_p &\mapsto c_p\big([\mathbf{c}_{jk}]_p\big) := \operatorname*{argmin}_{\mathbf{c}_{w,p} \in \mathcal{C}_p} d\big([\mathbf{c}_{jk}]_p,\mathbf{c}_{m,p}\big),
\end{align*}
\vspace{-0pt} 
where \( \mathcal{C}_p \) denotes the codebook for the \( p \)-th subspace, containing \( M \) centroids, and \( d(\cdot, \cdot) \) is a distance metric such as the Euclidean distance. The set \( \mathcal{C}_p = \{[\mathbf{c}_{jk}]_p \in \mathbb{R}^m \mid w_p([\mathbf{c}_{jk}]_p) = \mathbf{c_{m,p}}\} \) is referred to as the cluster for $m$-th centroid \( \mathbf{c}_{m,p} \), and $w_p$ is a quantizer to map $[\mathbf{c}_{jk}]_p$ to $\mathbf{c}_{m,p}$.

Each vector \( \mathbf{c}_{jk} \) can be represented by its closest centroid in the codebook. The centroid result of quantization mapping \( w(\cdot) \) is derived by training \( P \) codebooks using clustering algorithms, such as Lloyd's iteration~\cite{lloyd}. 


Moreover, to efficiently index and search class embedding vectors, we construct an inverted multi-index~\cite{inverted} using the Cartesian product of subspaces \( \mathcal{C} = \mathcal{C}_1 \times \mathcal{C}_2 \times \cdots \times \mathcal{C}_P\), where each subspace \( \mathcal{C}_p \) represents mapping results, the \( M \) centroids of the \( p \)-th subspace. The structure of the inverted multi-index can be represented:
\vspace{-5pt}
\begin{align*}
    &\small{\texttt{centroid vector}} \to \langle \small{\texttt{cluster IDs}} \rangle \\
    &\small{\texttt{cluster ID}} \to [\small{\texttt{centroid vector}}, \\
    &\quad \quad \quad \quad \quad \quad \langle \small{\texttt{patch vectors}} \rangle \, \small{\texttt{in cluster}}, \\
    &\quad \quad \quad \quad \quad \quad \langle \small{\texttt{bounding boxes}} \rangle \, \small{\texttt{of the vectors}}, \\
    &\quad \quad \quad \quad \quad \quad \langle \small{\texttt{patch IDs}} \rangle \, \small{\texttt{of the vectors}}, \\
    &\quad \quad \quad \quad \quad \quad \langle \small{\texttt{key frame IDs}} \rangle \, \small{\texttt{of the vectors}}]
\end{align*}

\vspace{-5pt}

In addition to indexing the embedding vectors, supplementary metadata such as key frame identifiers and bounding box coordinates are stored separately in a relational database. The two systems are linked through the shared \texttt{patch ID}, which serves as a unique key for each key frame. When a query retrieves candidate embeddings from the vector database, their corresponding \texttt{patch ID}s are used to fetch the relevant metadata from the relational database.

\subsection{Approximate Nearest Neighbor Search}
Approximate Nearest Neighbor Search (ANNS~\cite{anns}) facilitates efficient similarity-based querying in large-scale video datasets using an inverted multi-index structure.

Given a query vector \( \mathbf{q} \), the search begins by partitioning \( \mathbf{q} \) into \( P \) smaller components \( \mathbf{q} = \left([\mathbf{q}]_1, [\mathbf{q}]_2, \dots, [\mathbf{q}]_P\right)^\top \). The similarity between each component and its respective centroid is computed as \( s([\mathbf{q}]_p, \mathbf{c}_{m,p}) = [\mathbf{q}]_p \cdot \mathbf{c}_{m,p} \), where \( \mathbf{c}_{m,p} \in \mathcal{C}_p \) denotes a centroid in the codebook. Based on these similarity scores, the centroids are ranked, and the Top-\( A \) best clusters are chosen, forming a candidate cluster set \( \mathcal{S}_A \).

\newcommand{\INPUT}{\item[\textbf{Input:}]}
\newcommand{\OUTPUT}{\item[\textbf{Output:}]}
\begin{algorithm}[t]\small
\caption{Approximate Nearest Neighbor Search}
\label{algo:ann}
\begin{algorithmic}[1]
    \INPUT Query vector $\mathbf{q}$, Inverted Multi-Index $\mathcal{I}$, Number of clusters $M$, Number of cluster queried $A$, Number of neighbors $k$
    \OUTPUT Top-$k$ nearest neighbors
    
    \STATE Normalize and partition $\mathbf{q}$ into $\left([\mathbf{q}]_1, [\mathbf{q}]_2, \dots, [\mathbf{q}]_P\right)^\top$
    \FOR{each subspace $p = 1, \dots, P$}
        \FOR{each centroid $\mathbf{c}_{m,p} \in \mathcal{C}_p$}
            \STATE $s([\mathbf{q}]_p, \mathbf{c}_{m,p}) = [\mathbf{q}]_p \cdot \mathbf{c}_{m,p}$
        \ENDFOR
        \STATE $\mathcal{S}_A \gets \text{Top-}A(\{s([\mathbf{q}]_p, \mathbf{c}_{m,p}) \,|\, \mathbf{c}_{m,p} \in \mathcal{C}_p\})$
    \ENDFOR
    
    \FOR{each vector $[\mathbf{c}_a]_p \in \mathcal{S}_A$}
        \STATE Retrieve residual vector 
        $[\mathbf{r}_a]_p = [\mathbf{c}_a]_p - \mathbf{c}_{m,p}$
        \STATE $ s([\mathbf{q}]_p, [\mathbf{c}_a]_p) \approx s([\mathbf{q}]_p, \mathbf{c}_{m,p}) + [\mathbf{q}]_p \cdot [\mathbf{r}_a]_p$
    \ENDFOR
    
    \STATE $\mathcal{S}_A'\gets \text{Top-}k(\{s([\mathbf{q}]_p, [\mathbf{c}_a]_p) \,|\, [\mathbf{c}_a]_p \in \mathcal{S}_A\})$
    
    \FOR{each candidate $\mathbf{c}_a' \in \mathcal{S}_A'$}
        \STATE $s^{\text{exact}}(\mathbf{q}, \mathbf{c}_a') = \sum_{p=1}^{P} \left( [\mathbf{q}]_p \cdot [\mathbf{c}_a']_p \right)$
    \ENDFOR
    
    \STATE Determine patch ID $P^* = \arg\max_{P_j} \sum_{i=1}^{P} \mathbb{I}(P_i = P_j)$
    
    \STATE Sort top-$k$ candidates in descending order of $s^{\text{exact}}(\mathbf{q}, \mathbf{c}_a')$
    
    \STATE \textbf{Return} Top-$k$ candidate vectors with the highest scores and their final patch IDs.
\end{algorithmic}
\end{algorithm}
\vspace{-2pt}

Using the inverted multi-index, the candidate cluster set \( \mathcal{S}_A \) is employed to retrieve a list of potential nearest neighbors. For each vector \( [\mathbf{c}_a]_p \) in the candidate clusters, the approximate similarity score is calculated using \( s([\mathbf{q}]_p, [\mathbf{c}_a]_p) \approx s([\mathbf{q}]_p, \mathbf{c}_{m,p}) + [\mathbf{q}]_p \cdot [\mathbf{r}_a]_p \), where \( [\mathbf{r}_a]_p \) represents the residual vector of \( [\mathbf{c}_a]_p \) for its assigned centroid \( \mathbf{c}_{m,p} \), given by \( [\mathbf{r}_a]_p = [\mathbf{c}_a]_p - \mathbf{c}_{m,p} \). This residual is precomputed and stored in a distance lookup-table~\cite{quanti}. After obtaining the approximate similarity scores, we select the top-\( k \) candidates \( [\mathbf{c}_a']_p \in \mathcal{S}_A'\) in descending order. Finally, the exact similarity score for each of the top-\( k \) candidates is computed as \( s^{\text{exact}}(\mathbf{q}, \mathbf{c}_a) = \sum_{p=1}^{P} \left( [\mathbf{q}]_p \cdot [\mathbf{c}_a']_p \right) \). The candidates are re-ordered based on these exact scores. 

In cases where the selected candidate vectors are composed of parts originating from different database vectors. By counting the frequency of each patch ID within the components of a candidate vector, we select the patch ID with the highest occurrence as the final identifier~\cite{quanti}. Mathematically, let \( P_i \) denote the patch ID for the \( i \)-th component of a candidate vector, where \( i = 1, 2, \dots, P \). The final patch ID \( P^* \) is determined as follows: $ P^* = \arg\max_{P_j} \sum_{i=1}^{P} \mathbb{I}(P_i = P_j)$, where \( \mathbb{I}(P_i = P_j) \) is an indicator function that equals 1 if \( P_i = P_j \) and 0 otherwise. This equation ensures that the patch ID with the highest frequency among all components is selected as the final patch ID. The complete process is summarized in Algorithm~\ref{algo:ann}. Moreover, given a set of top-\( k \) nearest neighbors identified by the vector database, their bounding box coordinates and other contextual details can be seamlessly retrieved on the relational database.

\section{Query Strategy}\label{sec:query}
This section details fast search and cross-modality rerank.

\subsection{Top-\textit{k} Fast Search}

\begin{figure}[t]
    \begin{center}
     \includegraphics[width=0.9\linewidth]{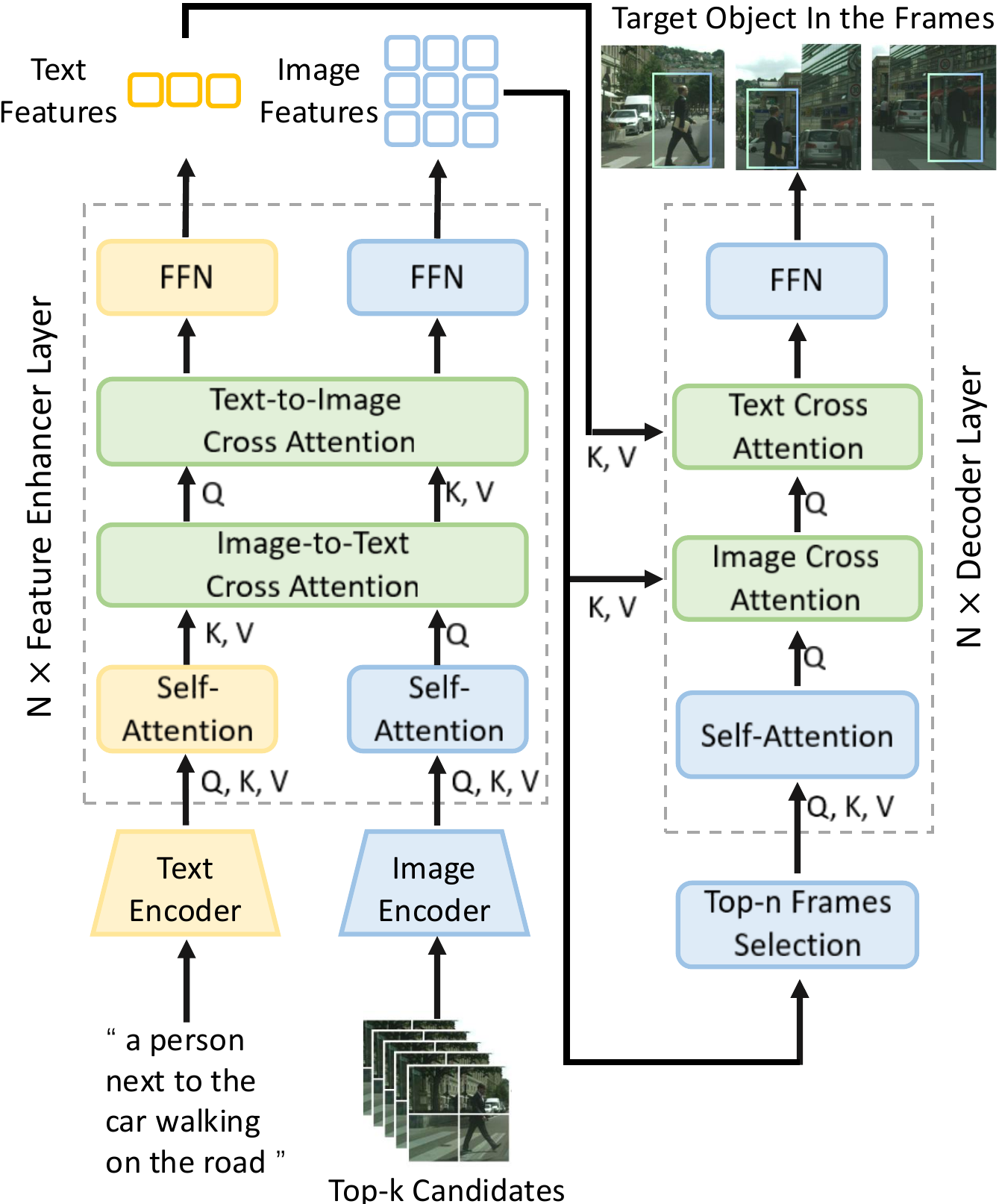}
     \vskip -0.1in
    \caption{Workflow of the cross-modality rerank. This leverages cross-modality cross-attention layers in feature enhancer and decoder to enhance modality alignment and applies Top-n frames rerank in Algorithm~\ref{algo:rerank}.}
    \label{fig:rerank}
    \end{center}
    \vspace{-10pt}
\end{figure}

To ensure efficient search, the user’s input text query is processed through a text encoder transformer to produce a semantic embedding vector. Notably, in the fast search stage, the entire sentence is encoded into a single feature vector, which reduces unnecessary associations with irrelevant details that may affect the downstream retrieval model. When input phrases contain multiple components, only the relevant phrases are extracted, discarding other words. This design removes cross-word dependencies and omits fine-grained positional information within the sentence. For instance, if the input text is “a person in black suit, walking on the road,” the encoder prioritizes extracting key phrases like “a person in black suit” and “road,” while ignoring finer relationships between them. This approach aligns well with the decoupled visual encoder, which emphasizes global object information and enables rapid retrieval of objects with distinctive features.

Using this text embedding, we search the large-scale vector database of video streams, computing similarity scores to quickly retrieve the top-\(k\) candidate patches corresponding to key frames. In the fast search phase, due to the independent operation of image and text encoders, the decoupled encoder setup is agnostic to any specific objects of interest that the user may have. Furthermore, by omitting intricate relationships (such as “walking on the road”), the text encoder’s representation is optimized for rapid preliminary retrieval. Following this, the top-\(k\) retrieved candidate frames undergo a more detailed feature extraction step. By leveraging the query text, this subsequent phase refines the retrieval process to yield a more accurate localization and identification of relevant objects within the candidate frames.

\subsection{Cross-Modality Rerank}

\begin{algorithm}[t]\small
\caption{Two-Stage Query Strategy}
\label{algo:rerank}
\begin{algorithmic}[1]
\INPUT Video collection \( V = \{v_i\}_{i=1}^M \), query text \( t \), fast search retrieval numbers \(k\), output frame numbers \(n\)

\OUTPUT Top-$n$ frames with bounding boxes of target objects
/*Stage 1: Top-\textit{k} Retrieval*/
\STATE Encode query text \( t \) into vector \( \mathbf{q} \)
\STATE Retrieve top-\(k\) candidates \(\{f_i\}_{i=1}^k\) using Algorithm~\ref{algo:ann}

/*Stage 2: Cross-Modality Rerank*/
\STATE $l_{s}$ $\gets$ \text{[ ]}
\FOR{each frame \( f_i \) in \(\{f_i\}_{i=1}^k\)}
    \STATE Extract enhanced features \( \mathbf{X}_I \in \mathbb{R}^{N_I \times d} \), \( \mathbf{X}_T \in \mathbb{R}^{N_T \times d} \)
    \[
    \mathbf\mathbf{X}_I, \mathbf\mathbf{X}_T \leftarrow \text{FeatureEnhancer}(\text{Encoder}(f_i, t))
    \]
    \STATE $l_{s} \leftarrow \max\limits_{j} \left( (X_I X_T^\top)_{j,-1} \right)$
\ENDFOR
\STATE Sort \(\{f_i\}_{i=1}^k\) in the descending order of $l_{s}$
\FOR{each feature $\mathbf{X}_I$ of Top-n frames}
    \STATE $\hat{b} \leftarrow Decoder(\mathbf{X}_I$, $\mathbf{X}_T)$
\ENDFOR
\STATE \textbf{Return:} Top-$n$ frames with bounding boxes $(f^*,\hat{b}^*)$
\end{algorithmic}
\end{algorithm}

As illustrated in Fig.~\ref{fig:rerank}, this stage is implemented using a cross-modality transformer model, which takes the query text and the top-k frames obtained from fast search stage as input. It reranks and re-scores the frames based on the query, outputting frames with bounding boxes that align with the query. The model extracts text features and image features using a general text encoder (e.g. BERT \cite{bert}) and image encoder (e.g. ViT \cite{vit}), respectively. The extracted features are fused and enhanced through a feature enhancer module\cite{glip, groundingdino}. Within this module, the image-to-text cross-attention layer uses $K_{text}$ and $V_{text}$ derived from text features and $Q_{image}$ from image features as the inputs to the attention layer, aligning image features with text features to capture query-relevant semantic information: $\text{Attention}(Q_{\text{image}}, K_{\text{text}}, V_{\text{text}}) = 
\text{softmax}\left(\frac{Q_{\text{image}} K_{\text{text}}^\top}{\sqrt{d_k}}\right) V_{\text{text}}$.
Conversely, the text-to-image cross-attention layer uses $K_{image}$ and $V_{image}$ derived from image features and $Q_{text}$ from text features as the inputs to enhance the visual understanding of the text features.

The output cross modality features from feature enhancer are fed into the cross-modality decoder\cite{groundingdino}, which shares a similar structure with the feature enhancer. It further achieves modality alignment through text and image cross-attention layers and outputs the frames with the bounding boxes. Algorithm~\ref{algo:rerank} outlines our two-stage retrieval process, where initial candidate selection via fast search is followed by a rerank using cross-attention to achieve precise object localization.

\section{Evaluation}\label{sec:eval}
In our evaluation experiments, we aim to answer the following research questions:
\begin{enumerate}
    \item How does LOVO perform on complex object query with existing methods?
    \item How efficient is LOVO in performing object query search in large-scale video?
    \item How scalable is LOVO across datasets of different sizes?
    \item What are the contributions of LOVO’s individual components to final accuracy and runtime?
\end{enumerate}

\subsection{Experiment Setup}

\textbf{System Configuration}. We run our experiments on a machine equipped with a single NVIDIA RTX 3090 GPU, 18 Intel Xeon Platinum 8352V CPU cores, and 32 GB of memory. The operating system is Ubuntu 20.04 with kernel version 5.15.0.
We use GPU for tasks such as feature encoding and cross-modality rerank, and we also employ the pre-trained ViT-B/32 visual encoder and text transformer. Our system has been integrated within the vector database, Milvus~\cite{milvus2}.

\begin{figure*}[t]
    \begin{center}
     \includegraphics[width=\textwidth]{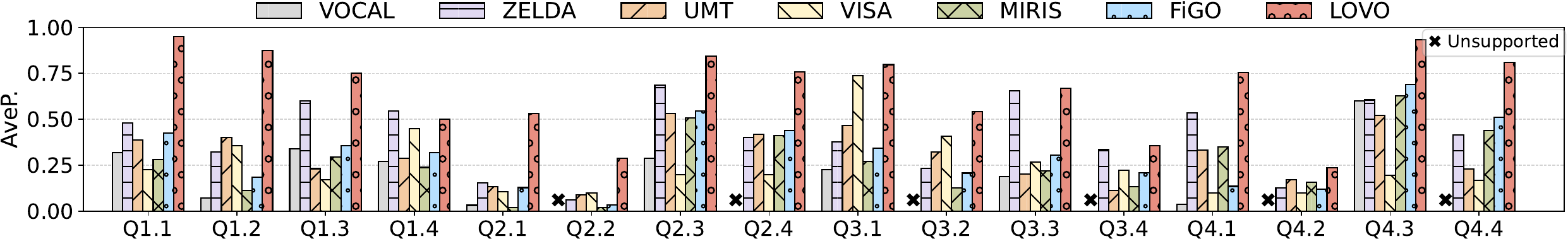}
     \vskip -0.08in
    \caption{Quantitative evaluation by average precision of LOVO against baselines.}
    \label{fig:exp1}
    \end{center}
    \vspace{-10pt}
\end{figure*}

\textbf{Datasets}. Table \ref{tab:queries} lists the datasets and queries for evaluation. To ensure a fair comparison of existing methods, we select the following representative datasets, widely adopted in various studies~\cite{focus,umt,miris,otif,leap}: 1) Cityscapes~\cite{city}: A video sequence dataset of urban traffic scenes from dashcams. We use the video in Stuttgart with size $59$GB. 2) Bellevue Traffic~\cite{bellevue}: A dataset of traffic surveillance videos from one intersections in Bellevue, with a total size of $62$GB and $101$ hours of recorded video. We utilized a $60$-minute subset for the experiment. 3) Qvhighlights~\cite{qvh}: A diverse dataset covering more than $10,000$ YouTube videos, each video has a duration of $150$ seconds. For the experiment, we selected $15$ videos from the QVHighlights evaluation dataset that matched the query scenarios. 4) Beach~\cite{miris}: A dataset containing 60 hours of video captured from a fixed camera on a resort’s sidewalk. We chose a $52$-minute video segment for the experiments. Bellevue and Beach datasets consist of fixed-camera footage, while Cityscapes and Qvhighlights involve moving cameras, making them more challenging. We used Bytetrack~\cite{zhang2022bytetrack} model to label bounding boxes for objects and manually checked and labeled the grounding truth frames for each query and dataset.

\textbf{Query Design}. 
To our knowledge, since there do not currently exist public any video datasets with complex object queries, we manually designed two sets of diverse object queries for each dataset, including both simpler object descriptions and more complex, detailed descriptions. These queries aligns with recent studies~\cite{otif, figo,equi,sketchql,zelda,leap}.

\renewcommand{\arraystretch}{1.2} 
\begin{table}[t!]
    \centering
    \caption{Example Queries for Different Datasets}
    \label{tab:queries}
    \begin{tabular}{c c p{4cm}}
    \hline
    \textbf{Dataset} & \textbf{Query ID} & \textbf{Query} \\ \hline
    \multirow{6}{*}{\textbf{Cityscapes}~\cite{city}}
        & Q1.1 & A person walking on the street. \\ \cline{2-3}
        & Q1.2 & A person in light-colored clothing walking while holding a dark bag. \\ \cline{2-3}
        & Q1.3 & A person riding a bicycle. \\ \cline{2-3}
        & Q1.4 & A person riding a bicycle, wearing a black t-shirt and blue jeans. \\ \hline
    \multirow{8}{*}{\textbf{Bellevue}~\cite{bellevue}}
        & Q2.1 & A red car driving in the center of the road. \\ \cline{2-3}
        & Q2.2 & A red car side by side with another car, both positioned in the center of the road. \\ \cline{2-3}
        & Q2.3 & A bus driving on the road. \\ \cline{2-3}
        & Q2.4 & A bus driving on the road with white roof and yellow-green body. \\ \hline
    \multirow{6}{*}{\textbf{Qvhighlights}~\cite{qvh}}
        & Q3.1 & A woman smiling sitting inside car. \\ \cline{2-3}
        & Q3.2 & A red-hair woman with white dress sitting inside a car. \\ \cline{2-3}
        & Q3.3 & A white dog inside a car. \\ \cline{2-3}
        & Q3.4 & A white dog inside a car, next to a woman wearing black clothes. \\ \hline
    \multirow{6}{*}{\textbf{Beach}~\cite{miris}}
        & Q4.1 & A green bus driving on the road. \\ \cline{2-3}
        & Q4.2 & A green bus with the white roof driving on the road. \\ \cline{2-3}
        & Q4.3 & A truck driving on the road. \\ \cline{2-3}
        & Q4.4 & A small white truck filled with cargo driving on the road. \\ \hline
    \end{tabular}
    \vspace{-10pt}
\end{table}

\textbf{Baselines}. We compare LOVO with several baselines across QA-index, QD-search, vision-based, and end-to-end methods.
\begin{itemize}
\item \textbf{VOCAL}~\cite{vexplore, equi, pvsg}. These QA-index methods employ spatio-temporal scene graphs to represent and index object relations within video frames.
\item \textbf{MIRIS}~\cite{miris}. This QD-search method aims for object tracking driven by the query. It is accelerated through offline object detector training and parameter tuning.
\item \textbf{FiGO}~\cite{figo}. FiGO is a QD-search method. It uses an ensemble of detection models to support a range of throughput-accuracy tradeoffs.
\item \textbf{ZELDA}~\cite{zelda}. As a vision-based method, ZELDA utilizes the vision-language model CLIP~\cite{clip}, enabling complex natural language queries for video frames. 
\item \textbf{UMT}~\cite{umt}. This end-to-end retrieval method, UMT, searches for events by employing temporal information, retrieving entire video moments rather than frames. 
\item \textbf{VISA}~\cite{visa}. As a video reasoning segmentation method, VISA leverages vision encoder and large language model to identify and segment objects across video frames.
\end{itemize}

\textbf{Metrics}. Our method is evaluated on the user-perceivable accuracy and query latency. We use the Average Precision (AveP) score as the performance metric, widely adopted in information retrieval tasks~\cite{zelda,seesaw}, defined as the area under the precision-recall curve: $\text{AveP} = \int_0^1 p(r) \, dr$, with \( p(r) \) representing precision at recall \( r \). We ranked each object retrieved in descending order based on the scores and calculated the precision and recall for each top-$n$. An object is considered a positive match when its intersection-over-union (IoU) exceeds 50\%, following the standard setting in object detection tasks (e.g. MSCOCO)~\cite{coco, zou2023object}. For each query, we compute the AveP by selecting the top-$10$ times of the grounding truth objects from each method, where the grounding truth number is the number of true positive samples. We evaluate the runtime efficiency by comparing the total execution time, including video processing and query search time.

\subsection{Query Accuracy}

\begin{figure}[t]
    \begin{center}
     \includegraphics[width=\linewidth]{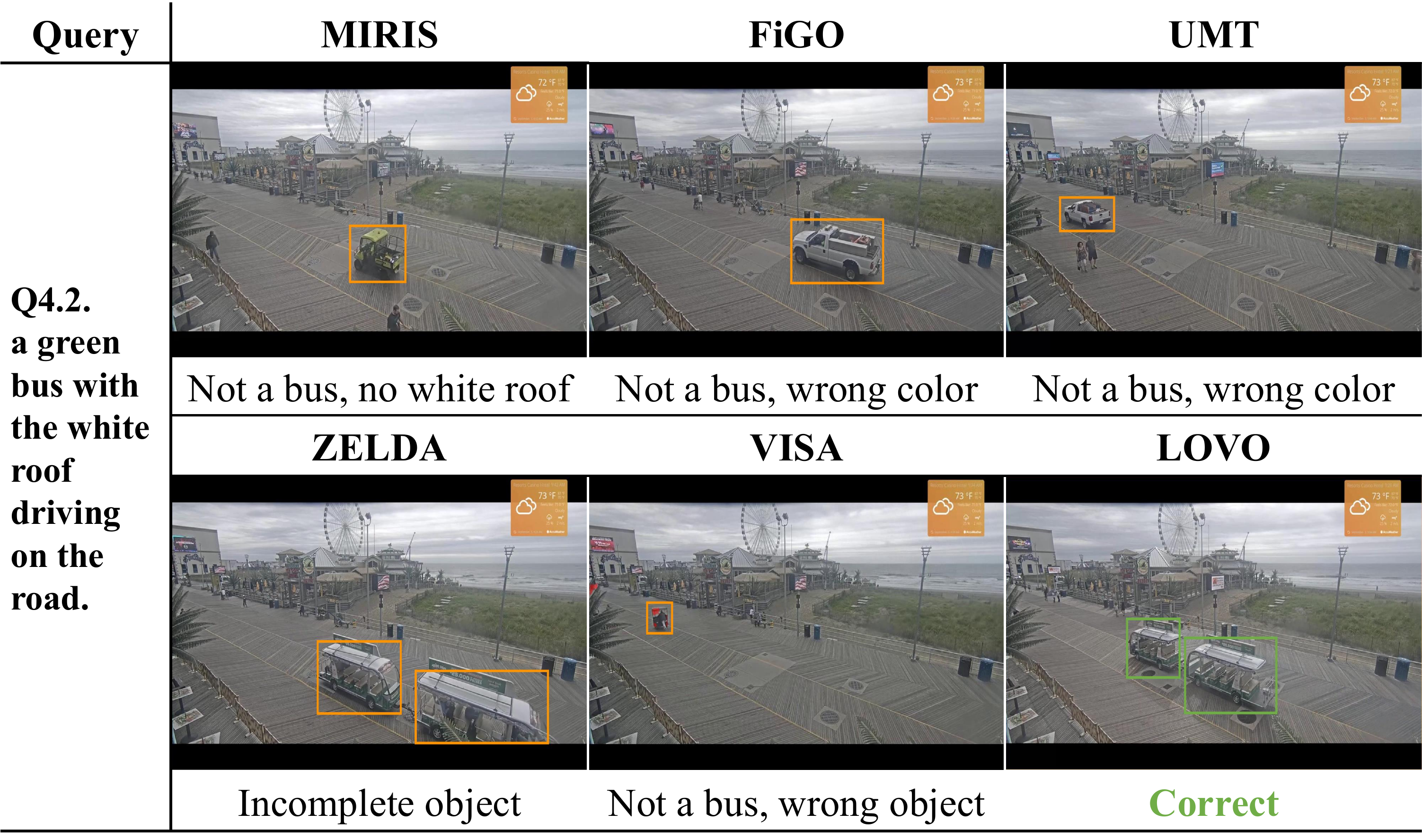}
     \vskip -0.05in
    \caption{Qualitative analysis of LOVO against  baselines. We here present representative frames in Q4.2 with the highest score retrieved by each baseline.} 
    \label{fig:exp2}
    \end{center}
    \vspace{-20pt}
\end{figure}
\textbf{Quantitative Analysis}. We measure the AveP of each method on various queries in Table~\ref{tab:queries}. As shown in Fig.~\ref{fig:exp1}, our method outperforms all baselines in AveP. This is largely attributed to the approximate indexing construction and cross-modality fusion, both improving the recall and precision of top-$n$ retrieval objects. We observe that VOCAL is nearly unable to recognize most of the queries since its index lacks the novel objects or location relations, like ``a red-hair woman" or ``side by side", so we omit it in the following experiments. MIRIS and FiGO exhibit lower precision due to the limited generality of their detection models and their reliance on manual, plan-based search mechanisms, which require additional model retraining to handle more complex queries. ZELDA performs well for global descriptions of objects but struggles with detailed context information. This may be due to the visual embeddings of target objects being relatively dispersed rather than clustered together and the wrong alignments of query text and visual features. UMT excels in moment retrieval but faces challenges when searching for small objects within frames. Besides, the training data for its model comes from everyday life scenes rather than road surveillance, and this bias somewhat limits its performance. VISA achieves high accuracy on the Qvhighlights datatset but performs poorly on the other traffic scenes datasets. A straightforward explanation is that VISA is more tailored to daily life scenarios, as reflected in its training datasets that consist of videos with high-quality annotations rather than footage from traffic cameras.

\textbf{Qualitative Analysis}. We execute textual queries and visually inspect all the retrieved objects in the corresponding key frames. The results are compared across baselines by displaying the top matching frame. As illustrated in Fig.~\ref{fig:exp2}, for Q4.2, LOVO successfully retrieved the correct object as the appointed grounding truth, whereas all other baselines exhibited certain issues with either the object class or detailed features. Despite the additional time spent configuring execution plans for MIRIS, its search results still failed to meet the requirements for ``white roof", while FiGO confused the query with a white bus. Zelda, which focuses on the global features of the frame, identified the largest but incomplete object partially satisfying the query. VISA incorrectly captured the objects as it was predominantly fine-tuned for moving objects segmentation within one video, rather than for object retrieval based on its detailed description.

\subsection{Runtime Efficiency}

\begin{figure}[h]
    \vskip -0.1in
    \centering
    \subfigure[Cityscapes.]{\includegraphics[width=0.23\textwidth]{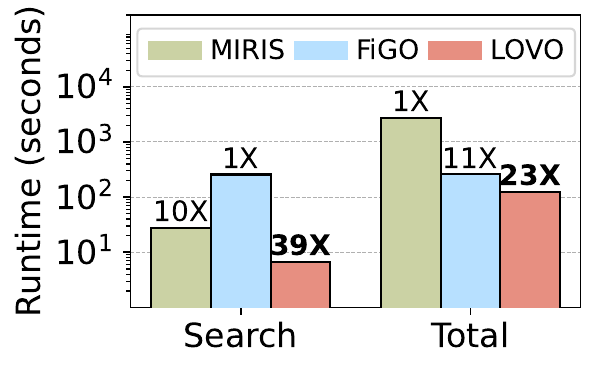} \label{fig:exp3_a}}
    \subfigure[Bellevue.]{\includegraphics[width=0.23\textwidth]{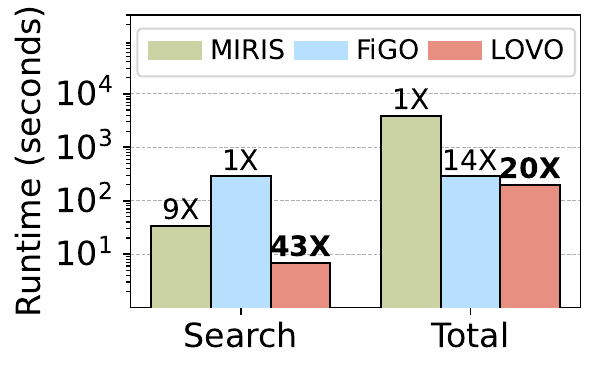} \label{fig:exp3_b}}
    \subfigure[Qvhighlights.]{\includegraphics[width=0.23\textwidth]{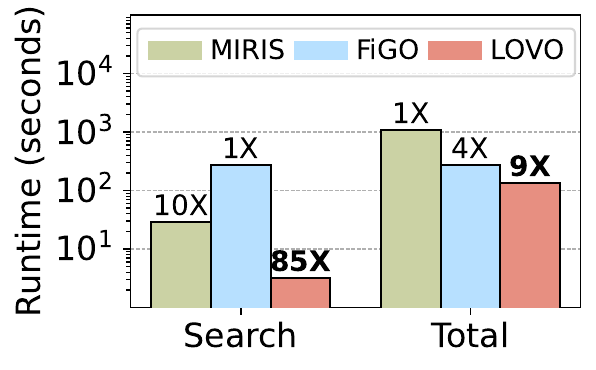}\label{fig:exp3_c}}
    \subfigure[Beach.]{\includegraphics[width=0.23\textwidth]{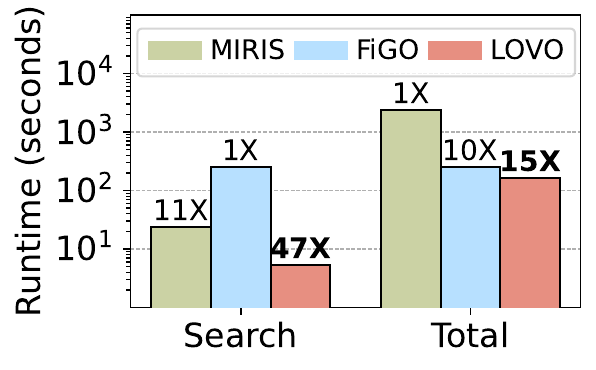}\label{fig:exp3_d}}
    \caption{Runtime comparison of MIRIS, FiGO, and LOVO across different datasets. We use the slowest method as the baseline and calculate the acceleration factors of the other two.}
    \label{fig:exp3_time}
    \vspace{-5pt}
\end{figure}

\textbf{Total Execution Time}. We measure the average total execution time for processing a single query across different datasets. Our method shows improvements in total execution time, with the highest acceleration rate of over $22$ times compared to MIRIS and over $2$ times compared to FiGO, as displayed in Fig.~\ref{fig:exp3_time}. On the one hand, MIRIS requires the most processing time across all four datasets due to the need for manual plan and model parameter adjustments. On the other hand, FiGO shows a slow performance for the additional query optimization time.

\textbf{Query Search Time}. The results in Fig.~\ref{fig:exp3_time} also demonstrate a significant improvement in average query search time for LOVO compared to MIRIS and FiGO, due to the optimized approximate indexing in a vector-based structure. LOVO achieves speeds at most $9$ times faster than MIRIS and $85$ times faster than FiGO. In particular, the application of ANNS in our system reduced the search space and improved query latency via fast similarity matching.

\textbf{Efficiency of Vision-Based and End-to-End Methods}. We compare the average time required for video processing and query search between LOVO, vision-based methods, and end-to-end method, as illustrated in Table \ref{tab:exp3_time}. These methods represent emerging techniques in the database domain and differ fundamentally from traditional index-based or search-based methods, so they are assessed separately. For query search time, our method significantly outperforms UMT, highlighting that UMT is primarily designed for moment retrieval tasks rather than for object queries on large-scale video datasets. Compared to ZELDA, LOVO has higher search times, despite incorporating a fine-grained cross-modality rerank stage. This additional step sacrifices some search time but substantially enhances overall accuracy in object queries. In addition, due to the increased computational burden brought by a larger number of parameters and the sequential token processing in LLM, both the processing and search time for VISA to execute queries is significantly higher than other methods.

\renewcommand{\arraystretch}{1.2}
\begin{table}[t]
    \centering
    \caption{Execution time (seconds) comparison of ZELDA, UMT, and LOVO. We note video processing, query search, and total execution time as Processing, Search, and Total in the table.}
    \label{tab:exp3_time}
    \begin{tabular}{p{1.22cm} l c c c c c}
    \hline
    \textbf{ } & \textbf{Phase} & \textbf{Cityscapes} & \textbf{Bellevue} & \textbf{Qvh.} & \textbf{Beach} \\ \hline
    \textbf{ZELDA}~\cite{zelda} & Processing & 141 & 215 & 141 & 56.5 \\ 
    & Search & \textbf{4.88} & \textbf{3.98} & 3.32 & \textbf{4.21} \\ 
    & Total & 146 & 218 & 145 & \textbf{60.7} \\ \hline
    \textbf{UMT}~\cite{umt} & Processing & \textbf{29.3} & \textbf{44.4} & \textbf{17.7} & \textbf{42.8} \\ 
    & Search & 104 & 122 & 54.7 & 93.8 \\ 
    & Total & \textbf{134} & \textbf{167} & \textbf{72.4} & 137 \\ \hline
    \textbf{VISA}~\cite{visa} & Processing & 326 & 613 & 744 & 316 \\ 
    & Search & 1564 & 430 & 346 & 194 \\ 
    & Total & 1890 & 1044 & 1090 & 510 \\  \hline
    \textbf{LOVO} & Processing & 118 & 192 & 117 & 155.2 \\ 
    & Search & 26.7 & 26.8 & 23.2 & 25.3 \\ 
    & Total & 145 & 220 & 152 & 185 \\ \hline
    \end{tabular}
\end{table}
\vspace{-10pt}

\begin{figure}[ht]
    \begin{center}
     \includegraphics[width=0.7\linewidth]{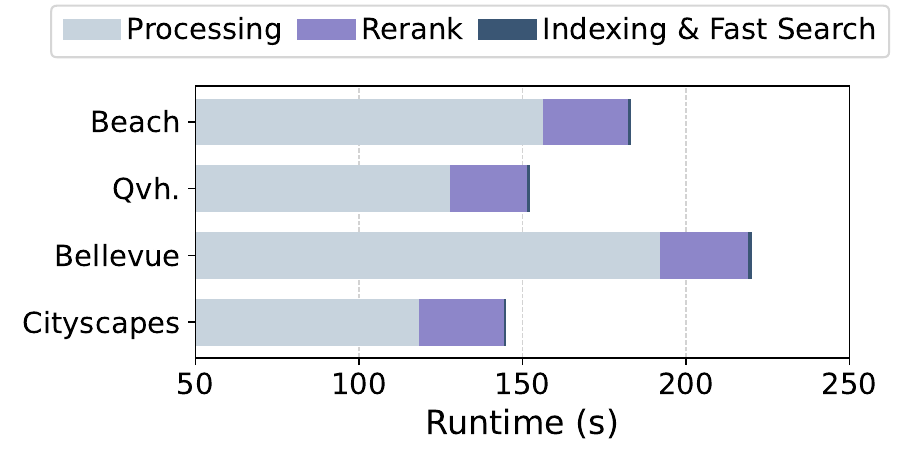}
     \vskip -0.1in
    \caption{Time distribution for query execution in different datasets.}
    \label{fig:pre5_dist}
    \end{center}
\end{figure}

\textbf{Model Time Distribution}. We further analyze the time distribution across the modules of our system, categorizing the total execution time into three phases. Given that the indexing and fast search phases are greatly short, we combine each other in comparison to the other phase. As depicted in Fig.~\ref{fig:pre5_dist}, LOVO effectively manages time allocation, with indexing and fast search being the fastest. Although the cross-modality rerank requires more time compared to the fast search, it still substantially reduces the need for repeatedly allocating multiple detection models to scan the entire videos like QD-search. Video processing, including generating visual embeddings, takes the longest but is performed offline, meaning users are not affected by this one-time computational overhead.

\begin{figure}[t]
    \centering
    \subfigure[Total execution time including processing, indexing, query search.]{\includegraphics[width=0.23\textwidth]{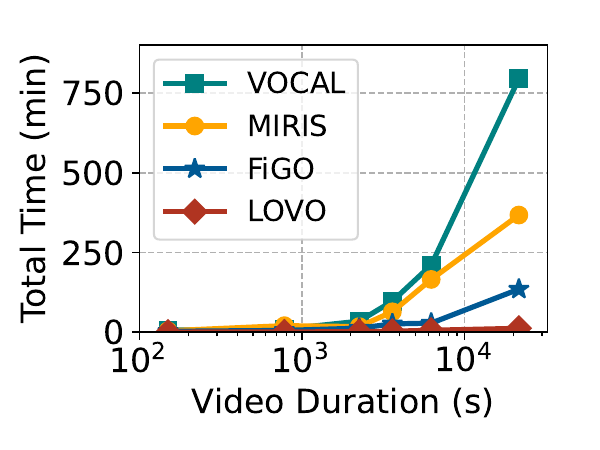} \label{fig:run_scale1}}
    \subfigure[Query search time perceived by users during query search.]{\includegraphics[width=0.23\textwidth]{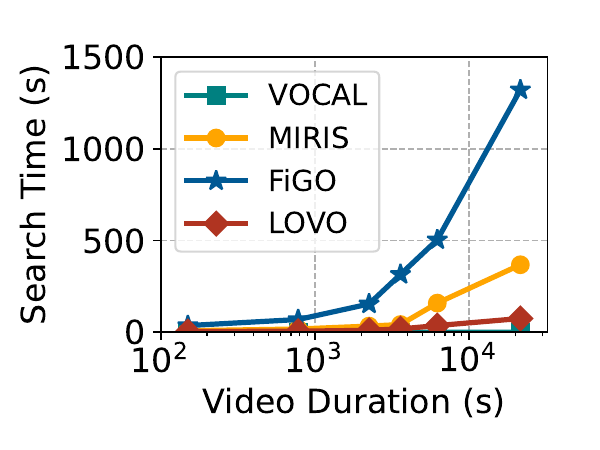} \label{fig:run_scale2}}
    \caption{Runtime comparison of total execution time and query search time.}
    \label{fig:runtime_compare}
    \vspace{-10pt}
\end{figure}

\subsection{Scalability Analysis}
\textbf{Dataset Scalability}. As shown in Fig.~\ref{fig:runtime_compare}, we evaluated the scalability of LOVO against existing methods such as VOCAL, FiGO, and MIRIS by measuring both the total execution time and the query search time as the duration of input video datasets increased. The results demonstrate that LOVO significantly outperforms the baselines in scalability. First, it performs one-time feature extraction and object detection during the video processing stage, avoiding redundant reprocessing for each query, unlike search-based methods that repeatedly process content. Second, LOVO integrates an inverted multi-index within a vector database, which allows efficient ANNS, thus facilitating rapid object recall even in extensive datasets. Third, LOVO applies cross-modality rerank on a small subset of candidate vectors, which ensures that rerank costs remain largely independent of the overall video size. VOCAL suffers from significant increases in indexing as the dataset size grows, or search-based methods MIRIS includes prolonged processing for detector training, FiGO repeatedly invokes multiple detection models for each query, resulting in significantly increased execution times.

\begin{figure}[t]
    \centering
    \subfigure[Scaling behavior across various frame numbers and processing time.]{\includegraphics[width=0.225\textwidth]{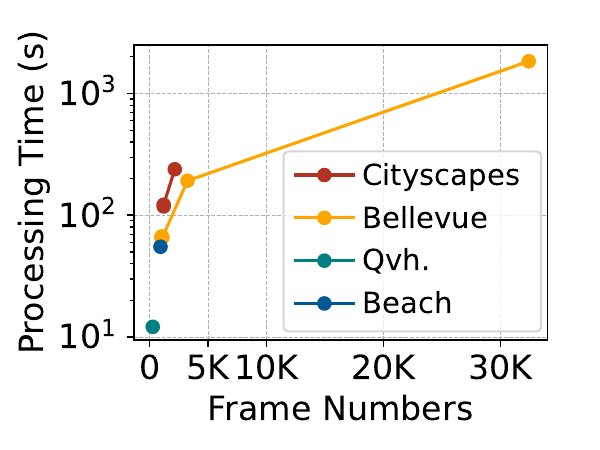}\label{fig:exp6_a}}  
    \subfigure[Scaling behavior across increasing indexing sizes and fast search time.]{\includegraphics[width=0.235\textwidth]{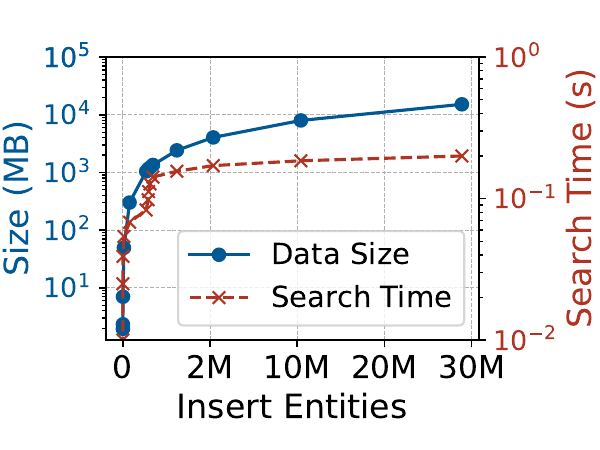}  \label{fig:exp6_b}}
    \subfigure[Distribution of fast search time per entity for each dataset.]
    {\includegraphics[width=0.23\textwidth]{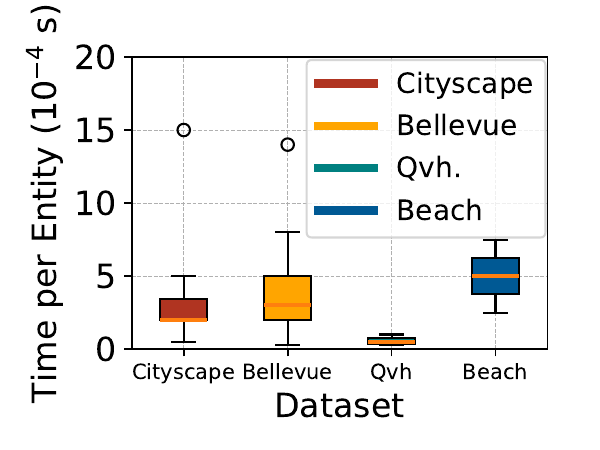}  \label{fig:exp6_d}}
    \subfigure[Cross-modality rerank time across different object numbers.]{\includegraphics[width=0.23\textwidth]{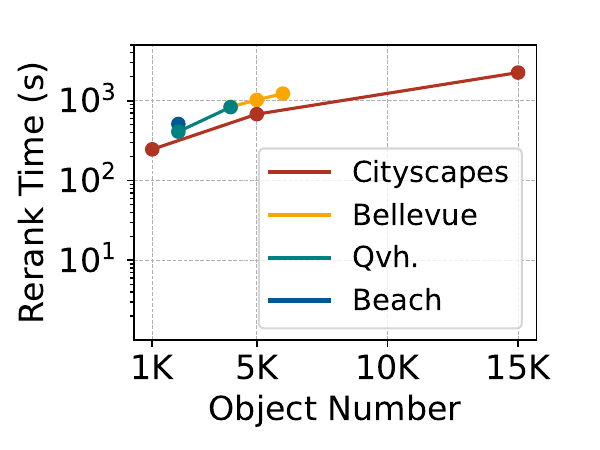} 
    \label{fig:exp6_c}}
    \caption{Comparison of execution times across different scales. }
    \label{fig:exp6_time}
\end{figure}

\textbf{Module Scalability}. Beyond comparing the different methods, we further analyze the scalability design among each individual component of LOVO. In Fig.~\ref{fig:exp6_a}, we measure its processing time for varying numbers of key frames. The processing time shows a linear relationship with the number of frames, with an estimated processing time of approximately $0.08$ seconds per frame. Fig.~\ref{fig:exp6_b} demonstrates that as the indexing size grows, the fast search time remains consistently low—well below $1$ second—indicating minimal impact from increased data volume due to the efficiency of the inverted indexing structure. In Fig.~\ref{fig:exp6_d}, the fast search time across different datasets consistently stays around $10^{-4}$ seconds per object, reflecting the efficiency of our system. Lastly, Fig.~\ref{fig:exp6_c} presents the time required for cross-modality rerank when querying different numbers of objects. The growth in rerank time required per query scales gradually as the number of objects increases, maintaining the capability to analyze one key frame in approximately $1$ second.

\subsection{Ablation Study}
\textbf{Impact of Cross-Modality Rerank}. We evaluate the influence of the cross-modality rerank stage on object query accuracy. The rerank process leverages cross-attention between visual and textual features to refine the initial results. The removal of the rerank stage results in a noticeable decline in overall accuracy, with object results becoming less relevant to the input queries. As shown in Table \ref{tab:ablation_combined}, In complex query types (Q2.2), the performance degradation caused by removing the rerank module is significantly greater than in relatively simple query types (Q1.1 and Q1.2).

\textbf{Impact of Approximate Nearest Neighbor Search}. We assess the role of ANNS in balancing query accuracy and system efficiency. Considering that the number of objects for rerank varies for each query, we also calculate the rerank time per frame as a metric. ANNS significantly reduces the search space by quickly identifying candidates. As shown in Table \ref{tab:ablation_combined}, the fast search time significantly increased after removing ANNS across four datasets, with an increase ranging from 57\% to 289\%, due to exhaustive searching. Additionally, the rerank time also experienced an increase. Although the use of ANNS involves an approximation that may introduce a slight compromise in recall (e.g. Q1.2 in Table \ref{tab:ablation_combined}), for the other three queries, the ANNS algorithm improved search accuracy, benefiting from its higher recall rate, proving the effectiveness of this approach for large-scale video datasets.

\textbf{Impact of Key Frame Selection.} As shown in Table~\ref{tab:ablation_combined}, LOVO reduces the number of processing frames through key frame selection, which helps reduce the fast query time (from $10^{-1}$ seconds to $10^{-2}$ seconds) and storage memory (from $7976$ MB to $2453$ MB, demonstrated in our experiment). Video frames often repeat similar content, with objects appearing across multiple frames. Extracting key frames avoids losing target objects, maintaining accuracy, while removing redundancy helps retrieve diverse objects from different parts of long videos, instead of focusing on one repeated object.

\renewcommand{\arraystretch}{1.2}
\begin{table}[t]
    \centering
    \caption{Ablation Study of LOVO on Cityscapes and Bellevue: Query Accuracy (AveP) and Latency (seconds). }
    \label{tab:ablation_combined}
    \begin{tabular}{l l c c c c}
    \hline
    \textbf{} & & \textbf{Q1.1} & \textbf{Q1.2} & \textbf{Q2.1} & \textbf{Q2.2} \\ \hline
    \multirow{3}{*}{\textbf{LOVO}} 
        & AveP          & \textbf{0.91} & 0.86          & 0.53 & \textbf{0.29} \\ 
        & Fast Search  & \textbf{0.06} & \textbf{0.09} & \textbf{0.03} & 0.07          \\  
        & Rerank       & \textbf{23.2} & 61.8 & \textbf{11.5} & \textbf{19.9} \\ \hline
        
    \multirow{3}{*}{\textbf{w/o Rerank}} 
        & AveP          & 0.80          & 0.75          & 0.44          & 0.09          \\ 
        & Fast Search  & 0.08          & \textbf{0.09} & \textbf{0.03} & \textbf{0.03} \\ 
        & Rerank       & --            & --            & --            & --            \\ \hline
    
    \multirow{3}{*}{\textbf{w/o ANNS}} 
        & AveP          & 0.80          & \textbf{0.90} & 0.49          & 0.23          \\ 
        & Fast Search  & 0.15          & 0.35          & 0.05          & 0.11          \\ 
        & Rerank       & 26.9          & 66.6          & 11.8          & 21.2          \\ \hline
    
    \multirow{3}{*}{\textbf{w/o Key frame}} 
        & AveP          & 0.90 & 0.88 & \textbf{0.58} & 0.28 \\ 
        & Fast Search  & 0.52 & 0.71 & 0.44 & 0.70 \\ 
        & Rerank       & 23.4 & \textbf{61.1} & 12.8 & 28.8 \\ \hline
        
    \end{tabular}
    
\end{table}

\subsection{Further Experiment}

\renewcommand{\arraystretch}{1.2}
\begin{table}[ht]
    \centering
    \caption{Query Accuracy and Latency (seconds) of LOVO across ANN variants  (BF: Brute-force search, IVF-PQ: Quantization-based inverted indexing, HNSW: Graph-based indexing).}
    \label{tab:ann_variant}
    \begin{tabular}{l l c c c c}
    \hline
    \textbf{} & & \textbf{Q1.1} & \textbf{Q1.2} & \textbf{Q1.3} & \textbf{Q1.4} \\ \hline
    \multirow{3}{*}{\textbf{LOVO(BF)}}
        & AveP    & 0.80 & \textbf{0.90} & \textbf{0.83} & 0.50 \\ 
        & Search & 27.05 & 66.79 & 27.34 & 89.47 \\ 
        & Total  & 277.31 & 317.05 & 277.60 & 339.38 \\ \hline
    \multirow{3}{*}{\textbf{LOVO(IVF-PQ)}} 
        & AveP    & \textbf{0.91} & 0.86 & 0.75 & 0.50 \\ 
        & Search & \textbf{23.80} & \textbf{62.70} & 24.92 & 90.12 \\ 
        & Total  & \textbf{260.42} & \textbf{299.32} & \textbf{261.54} & \textbf{326.74} \\ \hline
    \multirow{3}{*}{\textbf{LOVO(HNSW)}} 
        & AveP    & 0.80 & 0.88 & 0.78 & 0.50 \\ 
        & Search & 24.08 & 66.11 & \textbf{23.49} & \textbf{88.08} \\ 
        & Total  & 275.49 & 317.52 & 274.90 & 339.49 \\ \hline
    \end{tabular}
\end{table}

\textbf{ANN Variants. } We extended the LOVO by incorporating different ANN variants in Table~\ref{tab:ann_variant}, including brute-force search (BF), vector quantization-based inverted indexing (IVF-PQ), and graph-based indexing (HNSW). To address the challenges of large-scale video, LOVO(HNSW) achieves low latency in search performance, while LOVO(BF) delivers the highest accuracy but at the cost of higher latency. In the case of quantization-based indexing mentioned in the paper, LOVO(IVF-PQ) strikes a balance with low processing time and high-speed search comparable to graph-based methods. This is largely attributed to the fast vector processing enabled by quantization, making it particularly suitable for environments with limited memory resources.

\textbf{Query Types Extension.} To further demonstrate LOVO’s robustness and generalizability in handling diverse queries, we randomly selected 12 videos from ActivityNet-QA dataset \cite{actqa}. From its annotation, we randomly chose four yes/no questions as queries, shown in Table~\ref{tab:actqa}, using videos with a ``yes" answer as grounding truth. Given that baseline methods are not well-suited for question-answering-style retrieval, we focus here mainly on presenting LOVO’s results. Results in Table~\ref{tab:actqa_result} show that LOVO successfully finds the object in target videos and achieves promising performance.

\renewcommand{\arraystretch}{1.2}
\begin{table}[t!]
\centering
\caption{Extension queries from ActivityNet-QA.}
\label{tab:actqa}
\begin{tabular}{c c p{3.7cm}}
\hline
\textbf{Dataset} & \textbf{Query ID} & \textbf{Query}\\ \hline
    \multirow{6}{*}{\textbf{ActivityNet-QA}~\cite{actqa} }
    & EQ1 & does the car park on the meadow \\ \cline{2-3}
    & EQ2 & is the person with a hat a man \\ \cline{2-3}
    & EQ3 & is the person in the red life jacket outdoors \\ \cline{2-3}
    & EQ4 & is the person in a grey skirt dancing in the room \\ \hline
\end{tabular}
\vskip -0.1in
\end{table}
\vspace{-10pt}

\renewcommand{\arraystretch}{1.2}
\begin{table}[ht]
    \centering
    \caption{Query Accuracy and Latency (seconds) of LOVO in ActivityNet-QA Dataset.}
    \label{tab:actqa_result}
    \begin{tabular}{l l c c c c}
    \hline
    \textbf{} & & \textbf{EQ1} & \textbf{EQ2} & \textbf{EQ3} & \textbf{EQ4} \\ \hline
    \multirow{3}{*}{\textbf{LOVO}}
        & AveP    & 0.99 & 0.75 & 0.72 & 0.74 \\ 
        & Search & 127.92 & 131.09 & 130.61 & 130.90 \\ 
        & Total  & 187.09 & 190.26 & 189.78 & 190.07 \\ \hline
    \end{tabular}
\end{table}
\vspace{-10pt}
\section{Related Work}\label{sec:related}

\textbf{Query-Agnostic Index-Based Methods}. Indexing objects in large-scale video datasets to reduce query latency is a commonly used technique in video analysis~\cite{focus, otif, leap, tahoma, zilla, equi,vexplore, blazeit, exsample}. While this strategy reduces storage costs by indexing frequently appearing objects, less common objects may be missed, leading to degraded query performance. Solutions like difference detectors~\cite{noscope} and specialized neural network models ~\cite{equi, vexplore, focus, otif, tahoma, zilla, blazeit} attempt to reduce this gap but often introduce trade-offs between accuracy and efficiency. An alternative is to cluster video frames based on approximate object information at ingest time and use detection models during query search~\cite{focus,leap}. However, these indexes, including object classes, frame IDs, and limited spatial-temporal correlations, can hardly support complex object queries with detailed descriptions. 

\textbf{Query-Dependent Search-Based Methods}. In terms of query search, video systems~\cite{figo, noscope, miris, viva, PP, eva, clues, tvm} allow users to specify object and action classes as query predicates~\cite{otif} or apply spatial-temporal rules over frames~\cite{viva, clues}. They typically rely on pre-trained models, e.g., detectors, tracking models, or scene graph extraction models~\cite{pvsg}, to extract object information for query formulation. For example, SurvQ~\cite{survq} uses YOLO and transfer learning to identify predefined attributes and objects in surveillance video, and FemmIR~\cite{femmir} integrates multimodal data for cross-modal retrieval. However, they inevitably involve invoking multiple models to perform detection on the large volume of video frames during the query phase. Even worse, for complex queries beyond predefined classes, those systems need to select, train, and run other appropriate models to reprocess those massive frames, leading to significant latency and high computation redundancy.

\textbf{Vision-Language Models}. Current vision-language models have shown strong global retrieval capabilities using natural language queries. However, some models struggle with the global alignment between text and image~\cite{clip, clip4clip}, leading to poor performance in complex scenarios, particularly when recognizing small objects with fine-grained differences. Other large language model methods are generally suited for the event or moment analysis of shorter videos~\cite{umt,visa}, since they still require significant computational resources and result in extremely high latency.

\section{Conclusion}\label{sec:conclu}

We presented LOVO, a novel system designed to address the challenges of complex object queries in large-scale videos. By employing a one-time feature extraction approach, utilizing a vector database for efficient indexing, and combining approximate nearest neighbor searches with cross-modality rerank, LOVO overcomes the limitations of predefined classes, supporting detailed query descriptions while achieving low-latency, high-accuracy, and high-scalability. Experiments show that LOVO achieves the highest query accuracy while reducing search latency by up to 85 times, with latency remaining almost unaffected by data size. In summary, the architecture and techniques of LOVO establish it as an efficient solution for querying large-scale video datasets.

In the future, we aim to refine the vector database design by leveraging segmented parallel processing to reduce the overhead of full rebuilds during video updates and enhancing the incremental indexing strategy for new insertions, thereby improving adaptability while maintaining low latency and scalability. Besides, to enable fine-grained searches involving events, instances, and spatiotemporal relationships, we plan to extract those features during data preprocessing, develop sophisticated representations to capture object context, and integrate efficient indexing with event detection systems, enhancing LOVO’s utility across diverse scenarios. Furthermore, we plan to utilize large language models to annotate and describe video objects, constructing a larger-scale, more generalized dataset for a more effective assessment. 
\section*{Acknowledgment}

We thank anonymous reviewers for their insightful comments. This work has been partially supported by NSFC Grants 61932014, 62232011 and 62402315, the Shanghai Science and Technology Innovation Action Plan Grant 24BC3201200.

\bibliographystyle{IEEEtran}
\bibliography{IEEEabrv,ref}

\begin{thebibliography}{10}
\providecommand{\url}[1]{#1}
\csname url@samestyle\endcsname
\providecommand{\newblock}{\relax}
\providecommand{\bibinfo}[2]{#2}
\providecommand{\BIBentrySTDinterwordspacing}{\spaceskip=0pt\relax}
\providecommand{\BIBentryALTinterwordstretchfactor}{4}
\providecommand{\BIBentryALTinterwordspacing}{\spaceskip=\fontdimen2\font plus
\BIBentryALTinterwordstretchfactor\fontdimen3\font minus \fontdimen4\font\relax}
\providecommand{\BIBforeignlanguage}[2]{{%
\expandafter\ifx\csname l@#1\endcsname\relax
\typeout{** WARNING: IEEEtran.bst: No hyphenation pattern has been}%
\typeout{** loaded for the language `#1'. Using the pattern for}%
\typeout{** the default language instead.}%
\else
\language=\csname l@#1\endcsname
\fi
#2}}
\providecommand{\BIBdecl}{\relax}
\BIBdecl

\bibitem{cameras}
M.~Xu, T.~Xu, Y.~Liu, X.~Liu, G.~Huang, and F.~X. Lin, ``Supporting video queries on zero-streaming cameras,'' \emph{CoRR}, vol. abs/1904.12342, 2019.

\bibitem{focus}
K.~Hsieh, G.~Ananthanarayanan, P.~Bodik, S.~Venkataraman, P.~Bahl, M.~Philipose, P.~B. Gibbons, and O.~Mutlu, ``Focus: Querying large video datasets with low latency and low cost,'' in \emph{13th USENIX Symposium on Operating Systems Design and Implementation (OSDI 18)}.\hskip 1em plus 0.5em minus 0.4em\relax Carlsbad, CA: USENIX Association, Oct. 2018, pp. 269--286.

\bibitem{dovedb}
Z.~Xiao, D.~Zhang, Z.~Li, S.~Wu, K.~Tan, and G.~Chen, ``Dovedb: {A} declarative and low-latency video database,'' \emph{Proc. {VLDB} Endow.}, vol.~16, no.~12, pp. 3906--3909, 2023.

\bibitem{exsample}
O.~Moll, F.~Bastani, S.~Madden, M.~Stonebraker, V.~Gadepally, and T.~Kraska, ``Exsample: Efficient searches on video repositories through adaptive sampling,'' in \emph{2022 IEEE 38th International Conference on Data Engineering (ICDE)}.\hskip 1em plus 0.5em minus 0.4em\relax Kuala Lumpur, Malaysia: IEEE, 2022, pp. 2956--2968.

\bibitem{noscope}
D.~Kang, J.~Emmons, F.~Abuzaid, P.~Bailis, and M.~Zaharia, ``Noscope: optimizing neural network queries over video at scale,'' \emph{Proc. VLDB Endow.}, vol.~10, no.~11, p. 1586–1597, Aug. 2017.

\bibitem{number}
H.~Turtiainen, A.~Costin, T.~Lahtinen, L.~Sintonen, and T.~Hamalainen, ``Towards large-scale, automated, accurate detection of cctv camera objects using computer vision. applications and implications for privacy, safety, and cybersecurity.(preprint),'' \emph{arXiv preprint arXiv:2006.03870}, 2020.

\bibitem{zilla}
B.~Hu, P.~Guo, and W.~Hu, ``Video-zilla: An indexing layer for large-scale video analytics,'' in \emph{Proceedings of the 2022 International Conference on Management of Data}, ser. SIGMOD '22.\hskip 1em plus 0.5em minus 0.4em\relax New York, NY, USA: Association for Computing Machinery, 2022, p. 1905–1919.

\bibitem{trafficMonitor}
W.~Zhang, Q.~J. Wu, and H.~bing Yin, ``Moving vehicles detection based on adaptive motion histogram,'' \emph{Digital Signal Processing}, vol.~20, no.~3, pp. 793--805, 2010.

\bibitem{trafficControl}
M.~S. Shirazi and B.~T. Morris, ``Vision-based turning movement monitoring: count, speed \& waiting time estimation,'' \emph{IEEE Intelligent Transportation Systems Magazine}, vol.~8, no.~1, pp. 23--34, 2016.

\bibitem{prevention}
\BIBentryALTinterwordspacing
B.~C. Welsh, D.~P. Farrington, and S.~A. Taheri, ``Effectiveness and social costs of public area surveillance for crime prevention,'' \emph{Annual Review of Law and Social Science}, vol.~11, no. Volume 11, 2015, pp. 111--130, 2015. [Online]. Available: \url{https://www.annualreviews.org/content/journals/10.1146/annurev-lawsocsci-120814-121649}
\BIBentrySTDinterwordspacing

\bibitem{emergency}
S.-H. Kim and K.~Chung, ``Emergency situation monitoring service using context motion tracking of chronic disease patients,'' \emph{Cluster Computing}, vol.~18, pp. 747--759, 2015.

\bibitem{otif}
F.~Bastani and S.~Madden, ``{OTIF:} efficient tracker pre-processing over large video datasets,'' in \emph{{SIGMOD} '22: International Conference on Management of Data, Philadelphia, PA, USA, June 12 - 17, 2022}, Z.~G. Ives, A.~Bonifati, and A.~E. Abbadi, Eds.\hskip 1em plus 0.5em minus 0.4em\relax Philadelphia, PA, USA: {ACM}, 2022, pp. 2091--2104.

\bibitem{blazeit}
D.~Kang, P.~Bailis, and M.~Zaharia, ``Blazeit: Optimizing declarative aggregation and limit queries for neural network-based video analytics,'' \emph{Proc. {VLDB} Endow.}, vol.~13, no.~4, pp. 533--546, 2019.

\bibitem{seiden}
J.~Bang, G.~T. Kakkar, P.~Chunduri, S.~Mitra, and J.~Arulraj, ``Seiden: Revisiting query processing in video database systems,'' \emph{Proc. VLDB Endow.}, vol.~16, no.~9, p. 2289–2301, May 2023.

\bibitem{eva}
Z.~Xu, G.~T. Kakkar, J.~Arulraj, and U.~Ramachandran, ``Eva: A symbolic approach to accelerating exploratory video analytics with materialized views,'' in \emph{Proceedings of the 2022 International Conference on Management of Data}, ser. SIGMOD '22.\hskip 1em plus 0.5em minus 0.4em\relax New York, NY, USA: Association for Computing Machinery, 2022, p. 602–616.

\bibitem{everest}
Z.~Lai, C.~Han, C.~Liu, P.~Zhang, E.~Lo, and B.~Kao, ``Top-k deep video analytics: A probabilistic approach,'' in \emph{Proceedings of the 2021 International Conference on Management of Data}, ser. SIGMOD '21.\hskip 1em plus 0.5em minus 0.4em\relax New York, NY, USA: Association for Computing Machinery, 2021, p. 1037–1050.

\bibitem{figo}
J.~Cao, K.~Sarkar, R.~Hadidi, J.~Arulraj, and H.~Kim, ``Figo: Fine-grained query optimization in video analytics,'' in \emph{{SIGMOD} '22: International Conference on Management of Data, Philadelphia, PA, USA, June 12 - 17, 2022}, Z.~G. Ives, A.~Bonifati, and A.~E. Abbadi, Eds.\hskip 1em plus 0.5em minus 0.4em\relax Philadelphia, PA, USA: {ACM}, 2022, pp. 559--572.

\bibitem{tahoma}
M.~R. Anderson, M.~Cafarella, G.~Ros, and T.~F. Wenisch, ``Physical representation-based predicate optimization for a visual analytics database,'' in \emph{2019 IEEE 35th International Conference on Data Engineering (ICDE)}.\hskip 1em plus 0.5em minus 0.4em\relax Macau, China: IEEE, 2019, pp. 1466--1477.

\bibitem{leap}
Y.~Xu, D.~Zhang, S.~Zhang, S.~Wu, Z.~Feng, and G.~Chen, ``Predictive and near-optimal sampling for view materialization in video databases,'' \emph{Proc. ACM Manag. Data}, vol.~2, no.~1, Mar. 2024.

\bibitem{coco}
\BIBentryALTinterwordspacing
T.-Y. Lin, M.~Maire, S.~Belongie, L.~Bourdev, R.~Girshick, J.~Hays, P.~Perona, D.~Ramanan, C.~L. Zitnick, and P.~Dollár, ``Microsoft coco: Common objects in context,'' 2015. [Online]. Available: \url{https://arxiv.org/abs/1405.0312}
\BIBentrySTDinterwordspacing

\bibitem{equi}
E.~Zhang, M.~Daum, D.~He, B.~Haynes, R.~Krishna, and M.~Balazinska, ``Equi-vocal: Synthesizing queries for compositional video events from limited user interactions,'' \emph{Proc. VLDB Endow.}, vol.~16, no.~11, p. 2714–2727, Jul. 2023.

\bibitem{chen2022ranked}
Y.~Chen, X.~Yu, and N.~Koudas, ``Ranked window query retrieval over video repositories,'' in \emph{2022 IEEE 38th International Conference on Data Engineering (ICDE)}.\hskip 1em plus 0.5em minus 0.4em\relax IEEE, 2022, pp. 2776--2791.

\bibitem{PP}
Y.~Lu, A.~Chowdhery, S.~Kandula, and S.~Chaudhuri, ``Accelerating machine learning inference with probabilistic predicates,'' in \emph{Proceedings of the 2018 International Conference on Management of Data}, ser. SIGMOD '18.\hskip 1em plus 0.5em minus 0.4em\relax New York, NY, USA: Association for Computing Machinery, 2018, p. 1493–1508.

\bibitem{miris}
F.~Bastani, S.~He, A.~Balasingam, K.~Gopalakrishnan, M.~Alizadeh, H.~Balakrishnan, M.~Cafarella, T.~Kraska, and S.~Madden, ``Miris: Fast object track queries in video,'' in \emph{Proceedings of the 2020 ACM SIGMOD International Conference on Management of Data}, ser. SIGMOD '20.\hskip 1em plus 0.5em minus 0.4em\relax New York, NY, USA: Association for Computing Machinery, 2020, p. 1907–1921.

\bibitem{anderson2019physical}
M.~R. Anderson, M.~Cafarella, G.~Ros, and T.~F. Wenisch, ``Physical representation-based predicate optimization for a visual analytics database,'' in \emph{2019 IEEE 35th International Conference on Data Engineering (ICDE)}.\hskip 1em plus 0.5em minus 0.4em\relax IEEE, 2019, pp. 1466--1477.

\bibitem{dino}
S.~Liu, Z.~Zeng, T.~Ren, F.~Li, H.~Zhang, J.~Yang, Q.~Jiang, C.~Li, J.~Yang, H.~Su, J.~Zhu, and L.~Zhang, ``Grounding dino: Marrying dino with grounded pre-training for open-set object detection,'' 2024.

\bibitem{bellevue}
{City of Bellevue}, ``Traffic video dataset,'' \url{https://github.com/City-of-Bellevue/TrafficVideoDataset}, 2017, accessed: 2024-08-14.

\bibitem{mv}
L.~{Bommes}, X.~{Lin}, and J.~{Zhou}, ``Mvmed: Fast multi-object tracking in the compressed domain,'' in \emph{2020 15th IEEE Conference on Industrial Electronics and Applications (ICIEA)}, 2020, pp. 1419--1424.

\bibitem{owlvit}
M.~Minderer, A.~Gritsenko, A.~Stone, M.~Neumann, D.~Weissenborn, A.~Dosovitskiy, A.~Mahendran, A.~Arnab, M.~Dehghani, Z.~Shen, X.~Wang, X.~Zhai, T.~Kipf, and N.~Houlsby, ``Simple open-vocabulary object detection with vision transformers,'' \emph{ECCV}, 2022.

\bibitem{vit}
A.~Dosovitskiy, ``An image is worth 16x16 words: Transformers for image recognition at scale,'' \emph{arXiv preprint arXiv:2010.11929}, 2020.

\bibitem{quanti}
H.~Jegou, M.~Douze, and C.~Schmid, ``Product quantization for nearest neighbor search,'' \emph{IEEE transactions on pattern analysis and machine intelligence}, vol.~33, no.~1, pp. 117--128, 2010.

\bibitem{lloyd}
S.~Lloyd, ``Least squares quantization in pcm,'' \emph{IEEE Transactions on Information Theory}, vol.~28, no.~2, pp. 129--137, 1982.

\bibitem{inverted}
A.~Babenko and V.~Lempitsky, ``The inverted multi-index,'' in \emph{2012 IEEE Conference on Computer Vision and Pattern Recognition}, 2012, pp. 3069--3076.

\bibitem{anns}
P.~Wieschollek, O.~Wang, A.~Sorkine-Hornung, and H.~P.~A. Lensch, ``Efficient large-scale approximate nearest neighbor search on the gpu,'' in \emph{Proceedings of the IEEE Conference on Computer Vision and Pattern Recognition (CVPR)}, June 2016.

\bibitem{bert}
J.~Devlin, ``Bert: Pre-training of deep bidirectional transformers for language understanding,'' \emph{arXiv preprint arXiv:1810.04805}, 2018.

\bibitem{glip}
L.~H. Li, P.~Zhang, H.~Zhang, J.~Yang, C.~Li, Y.~Zhong, L.~Wang, L.~Yuan, L.~Zhang, J.-N. Hwang \emph{et~al.}, ``Grounded language-image pre-training,'' in \emph{Proceedings of the IEEE/CVF Conference on Computer Vision and Pattern Recognition}, 2022, pp. 10\,965--10\,975.

\bibitem{groundingdino}
\BIBentryALTinterwordspacing
S.~Liu, Z.~Zeng, T.~Ren, F.~Li, H.~Zhang, J.~Yang, Q.~Jiang, C.~Li, J.~Yang, H.~Su, J.~Zhu, and L.~Zhang, ``Grounding dino: Marrying dino with grounded pre-training for open-set object detection,'' 2024. [Online]. Available: \url{https://arxiv.org/abs/2303.05499}
\BIBentrySTDinterwordspacing

\bibitem{milvus2}
R.~Guo, X.~Luan, L.~Xiang, X.~Yan, X.~Yi, J.~Luo, Q.~Cheng, W.~Xu, J.~Luo, F.~Liu, Z.~Cao, Y.~Qiao, T.~Wang, B.~Tang, and C.~Xie, ``Manu: a cloud native vector database management system,'' \emph{Proceedings of the VLDB Endowment}, vol.~15, no.~12, pp. 3548--3561, 2022.

\bibitem{umt}
Y.~Liu, S.~Li, Y.~Wu, C.-W. Chen, Y.~Shan, and X.~Qie, ``Umt: Unified multi-modal transformers for joint video moment retrieval and highlight detection,'' in \emph{Proceedings of the IEEE/CVF Conference on Computer Vision and Pattern Recognition}.\hskip 1em plus 0.5em minus 0.4em\relax New Orleans, LA, USA: IEEE/CVF, 2022, pp. 3042--3051.

\bibitem{city}
M.~Cordts, M.~Omran, S.~Ramos, T.~Rehfeld, M.~Enzweiler, R.~Benenson, U.~Franke, S.~Roth, and B.~Schiele, ``The cityscapes dataset for semantic urban scene understanding,'' in \emph{Proc. of the IEEE Conference on Computer Vision and Pattern Recognition (CVPR)}, 2016.

\bibitem{qvh}
J.~Lei, T.~L. Berg, and M.~Bansal, ``Qvhighlights: Detecting moments and highlights in videos via natural language queries,'' 2021.

\bibitem{zhang2022bytetrack}
Y.~Zhang, P.~Sun, Y.~Jiang, D.~Yu, F.~Weng, Z.~Yuan, P.~Luo, W.~Liu, and X.~Wang, ``Bytetrack: Multi-object tracking by associating every detection box,'' in \emph{European conference on computer vision}.\hskip 1em plus 0.5em minus 0.4em\relax Springer, 2022, pp. 1--21.

\bibitem{sketchql}
R.~Wu, P.~Chunduri, A.~Payani, X.~Chu, J.~Arulraj, and K.~Rong, ``Sketchql: Video moment querying with a visual query interface,'' \emph{Proceedings of the ACM on Management of Data}, vol.~2, no.~4, pp. 1--27, 2024.

\bibitem{zelda}
F.~Romero, C.~Winston, J.~Hauswald, M.~Zaharia, and C.~Kozyrakis, ``Zelda: Video analytics using vision-language models,'' \emph{arXiv preprint arXiv:2305.03785}, 2023.

\bibitem{vexplore}
M.~Daum, E.~Zhang, D.~He, S.~Mussmann, B.~Haynes, R.~Krishna, and M.~Balazinska, ``Vocalexplore: Pay-as-you-go video data exploration and model building,'' \emph{Proc. VLDB Endow.}, vol.~16, no.~13, p. 4188–4201, Sep. 2023.

\bibitem{pvsg}
J.~Yang, W.~Peng, X.~Li, Z.~Guo, L.~Chen, B.~Li, Z.~Ma, K.~Zhou, W.~Zhang, C.~C. Loy, and Z.~Liu, ``Panoptic video scene graph generation,'' in \emph{Proceedings of the IEEE/CVF Conference on Computer Vision and Pattern Recognition (CVPR)}.\hskip 1em plus 0.5em minus 0.4em\relax Vancouver, BC, Canada: IEEE/CVF, June 2023, pp. 18\,675--18\,685.

\bibitem{clip}
A.~Radford, J.~W. Kim, C.~Hallacy, A.~Ramesh, G.~Goh, S.~Agarwal, G.~Sastry, A.~Askell, P.~Mishkin, J.~Clark, G.~Krueger, and I.~Sutskever, ``Learning transferable visual models from natural language supervision,'' in \emph{Proceedings of the 38th International Conference on Machine Learning}, ser. Proceedings of Machine Learning Research, M.~Meila and T.~Zhang, Eds., vol. 139.\hskip 1em plus 0.5em minus 0.4em\relax Virtual Event: PMLR, 18--24 Jul 2021, pp. 8748--8763.

\bibitem{visa}
Z.~Bai, T.~He, H.~Mei, P.~Wang, Z.~Gao, J.~Chen, Z.~Zhang, and M.~Z. Shou, ``One token to seg them all: Language instructed reasoning segmentation in videos,'' \emph{Advances in Neural Information Processing Systems}, vol.~37, pp. 6833--6859, 2025.

\bibitem{seesaw}
O.~Moll, M.~Favela, S.~Madden, V.~Gadepally, and M.~Cafarella, ``Seesaw: Interactive ad-hoc search over image databases,'' \emph{Proc. ACM Manag. Data}, vol.~1, no.~4, Dec. 2023.

\bibitem{zou2023object}
Z.~Zou, K.~Chen, Z.~Shi, Y.~Guo, and J.~Ye, ``Object detection in 20 years: A survey,'' \emph{Proceedings of the IEEE}, vol. 111, no.~3, pp. 257--276, 2023.

\bibitem{actqa}
Z.~Yu, D.~Xu, J.~Yu, T.~Yu, Z.~Zhao, Y.~Zhuang, and D.~Tao, ``Activitynet-qa: A dataset for understanding complex web videos via question answering,'' in \emph{AAAI}, 2019, pp. 9127--9134.

\bibitem{viva}
D.~Kang, F.~Romero, P.~Bailis, C.~Kozyrakis, and M.~Zaharia, ``Viva: An end-to-end system for interactive video analytics,'' in \emph{Proceedings of the 12th Conference on Innovative Data Systems Research (CIDR)}.\hskip 1em plus 0.5em minus 0.4em\relax Chaminade, CA, USA: www.cidrdb.org, January 2022.

\bibitem{clues}
D.~Chao, Y.~Chen, N.~Koudas, and X.~Yu, ``Optimizing video queries with declarative clues,'' \emph{Proc. VLDB Endow.}, vol.~17, no.~11, p. 3256–3268, Aug. 2024.

\bibitem{tvm}
T.~Zhong, Z.~Zhang, G.~Lu, Y.~Yuan, Y.-P. Wang, and G.~Wang, ``Tvm: A tile-based video management framework,'' \emph{Proceedings of the VLDB Endowment}, vol.~17, no.~4, pp. 671--684, 2023.

\bibitem{survq}
M.~Stonebraker, B.~Bhargava, M.~Cafarella, Z.~Collins, J.~McClellan, A.~Sipser, T.~Sun, A.~Nesen, K.~Solaiman, G.~Mani \emph{et~al.}, ``Surveillance video querying with a human-in-the-loop,'' in \emph{Proceedings of the Workshop on Human-In-the-Loop Data Analytics with SIGMOD}, 2020.

\bibitem{femmir}
K.~Solaiman and B.~Bhargava, ``Feature centric multi-modal information retrieval in open world environment (femmir),'' \emph{Authorea Preprints}, 2023.

\bibitem{clip4clip}
H.~Luo, L.~Ji, M.~Zhong, Y.~Chen, W.~Lei, N.~Duan, and T.~Li, ``Clip4clip: An empirical study of clip for end to end video clip retrieval and captioning,'' \emph{Neurocomputing}, vol. 508, pp. 293--304, 2022.

\end{thebibliography}




\end{document}